\documentclass[twocolumn]{aastex63}

\usepackage{csvsimple}
\usepackage{gensymb}
\usepackage{dcolumn}
\usepackage{graphicx}
\usepackage{amsmath}
\usepackage{mathtools}
\usepackage{wrapfig}
\usepackage{multirow}
\usepackage{comment}
\usepackage{changepage}
\usepackage{lipsum}
\usepackage{xcolor}

\usepackage{array}

\graphicspath{{./}{figure*s/}}

\shorttitle{Chemo-Dynamically Tagged Groups of Halo $r$-Process-Enhanced Stars}
\shortauthors{Gudin et al.}

\begin{document}

\title{The \emph{R}-Process Alliance: Chemo-Dynamically Tagged Groups of Halo $r$-Process-Enhanced Stars Reveal a Shared Chemical-Evolution History}

\author{Dmitrii Gudin}
\affiliation{Department of Physics, University of Notre Dame, Notre Dame, IN 46556, USA}
\affiliation{Joint Institute for Nuclear Astrophysics -- Center for the Evolution of the Elements (JINA-CEE), USA}
\affiliation{Department of Mathematics, University of Maryland, College Park, MD 20742, USA}

\author{Derek Shank}
\affiliation{Department of Physics, University of Notre Dame, Notre Dame, IN 46556, USA}
\affiliation{Joint Institute for Nuclear Astrophysics -- Center for the Evolution of the Elements (JINA-CEE), USA}

\author{Timothy C. Beers}
\affiliation{Department of Physics, University of Notre Dame, Notre Dame, IN 46556, USA}
\affiliation{Joint Institute for Nuclear Astrophysics -- Center for the Evolution of the Elements (JINA-CEE), USA}

\author{Zhen Yuan}
\affiliation{Key Laboratory for Research in Galaxies and Cosmology,
Shanghai Astronomical Observatory, Chinese Academy of
Sciences, 80 Nandan Road, Shanghai 200030, China}

\author{Guilherme Limberg}
\affil{Universidade de S\~ao Paulo, Instituto de Astronomia, Geof\'isica e Ci\^encias Atmosf\'ericas, Departamento de Astronomia, SP 05508-090, S\~ao Paulo, Brazil}

\author{Ian U. Roederer}
\affiliation{Department of Astronomy, University of Michigan, 1085 S. University Ave., Ann Arbor, MI 48109, USA}
\affiliation{Joint Institute for Nuclear Astrophysics -- Center for the Evolution of the Elements (JINA-CEE), USA}

\author{Vinicius Placco}
\affiliation{Department of Physics, University of Notre Dame, Notre Dame, IN 46556, USA}
\affiliation{Joint Institute for Nuclear Astrophysics -- Center for the Evolution of the Elements (JINA-CEE), USA}
\affiliation{NSF's National Optical-Infrared Astronomy Research Laboratory, Tucson, AZ 85719, USA}

\author{Erika M.\ Holmbeck}
\affiliation{Department of Physics, University of Notre Dame, Notre Dame, IN 46556, USA}
\affiliation{Joint Institute for Nuclear Astrophysics -- Center for the Evolution of the Elements (JINA-CEE), USA}
\affiliation{Center for Computational Relativity and Gravitation, Rochester Institute of Technology, Rochester, NY 14623, USA} 

\author{Sarah Dietz}
\affiliation{Department of Physics, University of Notre Dame, Notre Dame, IN 46556, USA}
\affiliation{Joint Institute for Nuclear Astrophysics -- Center for the Evolution of the Elements (JINA-CEE), USA}

\author{Kaitlin C. Rasmussen}
\affiliation{Department of Physics, University of Notre Dame, Notre Dame, IN 46556, USA}
\affiliation{Joint Institute for Nuclear Astrophysics -- Center for the Evolution of the Elements (JINA-CEE), USA}
\affiliation{Department of Astronomy, University of Michigan, 1085 S. University Ave., Ann Arbor, MI 48109, USA}

\author{Terese T.\ Hansen}
\affiliation{George P.~and Cynthia Woods Mitchell Institute for Fundamental Physics and Astronomy, Texas A\&M University, College Station, TX 77843, USA}
\affiliation{Department of Physics and Astronomy, Texas A\&M University, College Station, TX 77843, USA}

\author{Charli M.\ Sakari}
\affiliation{Department of Physics and Astronomy, San Francisco State University, San Francisco, CA 94132, USA}

\author{Rana Ezzeddine}
\affiliation{Department of Physics and Kavli Institute for Astrophysics and Space Research, Massachusetts Institute of Technology, Cambridge, MA 02139, USA}
\affiliation{Joint Institute for Nuclear Astrophysics -- Center for the Evolution of the Elements (JINA-CEE), USA}
\affiliation{Department of Astronomy, University of Florida, Bryant Space Science Center, Gainesville, FL 32611, USA}

\author{Anna Frebel}
\affiliation{Department of Physics and Kavli Institute for Astrophysics and Space Research, Massachusetts Institute of Technology, Cambridge, MA 02139, USA}
\affiliation{Joint Institute for Nuclear Astrophysics -- Center for the Evolution of the Elements (JINA-CEE), USA}

\begin{abstract}
We derive dynamical parameters for a large sample of 446 $r$-process-enhanced (RPE) metal-poor stars in the halo and disk systems of the Milky Way, based on data releases from the $R$-Process Alliance, supplemented by additional literature samples.  This sample represents more than a ten-fold increase in size relative to that previously considered by Roederer et al., and, by design, covers a larger range of $r$-process-element enrichment levels.  We test a number of clustering analysis methods on the derived orbital energies and other dynamical parameters for this sample, ultimately deciding on application of the \texttt{HDBSCAN} algorithm, which obtains 30 individual Chemo-Dynamically Tagged Groups (CDTGs); 21 contain between 3 and 5 stars, and 9 contain between 6 and 12 stars. Even though the clustering was performed solely on the basis of their dynamical properties, the stars in these CDTGs exhibit {\it statistically significant similarities} in their metallicity ([Fe/H]), carbonicity ([C/Fe]), and neutron-capture element ratios ([Sr/Fe], [Ba/Fe], and [Eu/Fe]). These results demonstrate that the RPE stars in these CDTGs have likely experienced common chemical-evolution histories, presumably in their parent satellite galaxies or globular clusters, prior to being disrupted into the Milky Way's halo. We also confirm the previous claim that the orbits of the RPE stars preferentially exhibit pericentric distances that are substantially lower than the present distances of surviving ultra-faint dwarf and canonical dwarf spheroidal galaxies, consistent with the disruption hypothesis. The derived dynamical parameters for several of our CDTGs indicate their association with previously known substructures, Dynamically Tagged Groups, and RPE Groups.
\end{abstract}


\section{Introduction} 

The origin of the heaviest elements ($Z\gtrsim 30$) found in our Milky Way has been debated for over half a century, beginning with the pivotal works of \citet{burbidge1957} and \citet{cameron1957}. These elements, the great majority of which were formed by neutron-capture processes, have been measured for numerous individual stars in the halo and disk populations of the Galaxy, as well as in dwarf satellite galaxies and globular clusters.  

The astrophysical site(s) of the slow neutron-capture process ($s$-process) are reasonably well understood -- the thermal pulses of asymptotic giant-branch stars (in some cases preserved on an observed companion due to subsequent mass transfer across a binary system; see, e.g., \citealt{herwig2005, bisterzo2010, abate2015}), or in massive, rapidly rotating, extremely low-metallicity stars (e.g., \citealt{meynet2006d}; \citealt{pignatari2008}; \citealt{choplin2018}).
Recent work has provided evidence for the likely operation of the so-called intermediate neutron-capture process ($i$-process; \citealt{cowan1977}), the astrophysical site(s) for which are still under study \citep{dardelet2014,hampel2016,denissenkov2017}. Heavy elements that were produced by the rapid neutron-capture process ($r$-process) have had multiple proposed astrophysical sites (e.g., \citealt{arnould2007}; \citealt{frebel2018}), including neutron star mergers (NSMs; \citealt{lattimer1974,thielemann2017}), magneto-rotational supernovae jets \citep{winteler2012, nishimura2015},
core-collapse supernovae (CCSNe; \citealt{arcones2007, wanajo2013}), and collapsars \citep{siegel2019}.

\begin{deluxetable*}{lc}
\tablecaption{RPA $r$-Process-Enhanced Star Classifications\label{tab1}}
\tablehead{\colhead{Sub-class} & \colhead{Abundances}}
\startdata
$r$-II &  [Eu/Fe] $>$ +0.7, [Ba/Eu] $<$ 0\\
$r$-I & +0.3 $<$ [Eu/Fe] $\leq$ +0.7, [Ba/Eu] $<$ 0\\
limited-$r$ & [Sr/Ba] $>$ +0.5, [Eu/Fe] $\leq +0.3$\\
CEMP-$r$ & [C/Fe] $> $ +0.7, [Eu/Fe] $>$ +0.3, [Ba/Eu] $<$ 0
\enddata 
\end{deluxetable*}

Recent observations of highly $r$-process-enhanced (hereafter, RPE) stars in the satellite ultra-faint dwarf (UFD) galaxy Reticulum II \citep{ji2016,ji2019,roederer2016}, coupled with the gravitational-wave detection of a NSM by LIGO/VIRGO (GW 170817; \citealt{abbott2017}), with subsequent photometric and spectroscopic analysis of its associated kilonova (SSS17a; e.g.,  \citealt{drout2017,kilpatrick2017,shappee2017,watson2019}), provided strong evidence for two key pieces of the puzzle for the astrophysical origin of the $r$-process elements. First, that NSMs can plausibly be the dominant source of the heavy ($Z>52$) $r$-process elements (e.g., \citealt{hotokezaka2018,horowitz2019}) -- although it is important to note that theoretical predictions of the NSM yields and frequencies vary, leading to significant uncertainties in the predicted amounts of expected $r$-process elements produced (e.g., \citealt{cote2019,keegans2019,siegel2019}).  Secondly, that dwarf spheroidal (dSph) galaxies and  UFDs may be the primary birth places of the metal-poor RPE stars in the halo of the Milky Way, which were later distributed throughout it when their parent dwarfs were disrupted during accretion (see, e.g., \citealt{brauer2019})\footnote{Globular clusters also possess RPE stars (e.g., \citealt{sneden2000,sobeck2011}), and since they may be accreted by the halo as well, could contribute a number of RPE stars to the halo system.  \citet{sakari2018a} identify the RPE star J2116-0213 as one possible example, based on its distinctive elevated [Na/Fe] and low [Mg/Fe] abundances.}. 

This opens the exciting possibility of grouping RPE stars on the basis of the similarity in their dynamical properties, which are expected to be approximately conserved even after their parent galaxy (or globular cluster) is destroyed (see, e.g., \citealt{roederer2018a}).  Alternatively, identification of Dynamically Tagged Groups (DTGs) in large catalogs of very metal-poor (VMP; [Fe/H] $< -2.0$) stars, followed by ``mapping" of the derived dynamical parameters of individual RPE stars onto the identified DTGs, provides another means to investigate their likely progenitor environments (see, e.g., \citealt{yuan2020} and \citealt{limberg2020}). We distinguish between these approaches by noting that the stellar DTGs are identified without knowledge of the detailed chemistry of their individual members, while the identification of dynamical groups of {\it previously known} chemically peculiar stars (such as RPE stars or carbon-enhanced metal-poor, CEMP,  stars) produces Chemo-Dynamically Tagged Groups (CDTGs). 

Application of mathematical clustering algorithms to the space of orbital energies and other suitable dynamical parameters will result in stars being grouped based on the similarity of their orbits around the Galaxy. Although such clustering is made without relying on knowledge of their chemistry, if the stars in a given CDTG indeed originated in a common progenitor galaxy or globular cluster, then they might be expected to exhibit smaller dispersions in their metallicities and other chemical species, including $r$-process-element abundances, than expected by chance. In the work of \cite{roederer2018a}, based on a relatively small sample of 35 RPE stars judged to have suitably accurate estimates of distances (out of an initial sample of 83 RPE stars), the metallicity spread for member stars in their individual RPE Groups was indeed shown to be smaller than expected for random draws from the full sample. The mean metallicities of their eight RPE Groups indicated likely association with low-mass dwarf galaxy parents, based on the metallicity-luminosity relationship \citep{kirby2013}. We note that tests on the similarity of the reported [Eu/Fe] ratios are somewhat compromised for this sample, since they are relatively few in number, and the range of the [Eu/Fe] ratios covered by their stars ([Eu/Fe] $> +0.7$) does not include the moderately $r$-process-enhanced $r$-I stars with $+0.3 < \rm [Eu/Fe] \leq +0.7$.  We address this difficulty in the present paper, based on consideration of a significantly larger sample of RPE stars, including the known $r$-I stars, spanning a wider range of [Eu/Fe].

The RPE stars can be split into multiple sub-classes, based on their light-element and neutron-capture-element abundances (see Table \ref{tab1}). For the purpose of the current effort, we make use of the classifications employed by the $R$-Process Alliance (RPA), which are a subset of the original classifications from \cite{beers2005}, with an extension from \citet{frebel2018}.  Note that the recent work of \citet{holmbeck2020} has advocated, from a statistical analysis of the complete set of RPA abundances measured to date, that the most appropriate division between the moderately enhanced $r$-I stars and the highly enhanced $r$-II stars is at [Eu/Fe] $> +0.7$ (rather than [Eu/Fe] $> +1.0$ in common previous use), which we have adopted here.

The $r$-II and $r$-I stars  are thought to have been enriched primarily by the main $r$-process.  Some $r$-II and  $r$-I stars exhibit enhanced abundances of the actinide elements, such as uranium and thorium \citep{hill2002, mashonkina2014, holmbeck2018}. The origin of this ``actinide boost'' \citep{schatz2002} is not known with certainty, but its presence hints at the possibility of fundamental differences in the nature of the $r$-process \citep{holmbeck2019b},, even within a single progenitor population (see, e.g., \citealt{holmbeck2019a}). There are insufficient numbers of actinide-boost stars known to confidently assign them into a separate class at present. The limited-$r$ stars exhibit enhanced light $r$-process elements (e.g., Sr, Y, Zr) relative to heavy ones (e.g., Ba, Eu), while the heavier elements are not enhanced \citep{frebel2018}\footnote{We point out that the limited-$r$ process has been recognized, under various names, over the past few decades.  See, e.g., \citet{McWilliam1998}, \citet{Travaglio2004}, and \citet{Hansen2012}, and references therein.} . These stars may have received significant contributions from CCSNe. The CEMP-$r$ stars exhibit carbon enhancement ([C/Fe] $> +0.7$), along with $r$-process-element enrichment (see, e.g., \citealt{beers2005}).   

In this work, we consider the dynamical properties of a large sample of 446 $r$-II and $r$-I stars from the RPA and other literature sources. This paper is arranged as follows. Section 2 describes the assembly of our sample, their derived distance estimates, and the available chemical-abundance information.  In Section 3, we derive orbital energies and other dynamical parameters for each of these stars, making use of parallaxes, proper motions, and radial velocities from \textit{Gaia} Data Release 2 (DR2) \citep{gaia2016,gaia2018a,gaia2018b,gaia2018c}, supplemented by kinematic data available in the literature, as needed. Section 4 describes the application of the \texttt{HDBSCAN} \citep{mcinnes2017} clustering algorithm to identify 30 CDTGs, populated with between 3 and 12 stars each. In Section 5, we investigate the distributions of various elemental abundances within each CDTG, and demonstrate that their dispersions are significantly smaller and globally inconsistent with random draws from the full population of RPEs.  Section 6 provides a discussion of the possible associations of our identified CDTGs with previously known substructures and DTGs, as well as how the pericentric distances of our RPEs (and CDTGs) compare with those of surviving dSph and UFD galaxies, and the implications of our CDTG chemistry on the nature of their birth environments and $r$-process progenitor(s). Section 7 summarizes our results, and provides perspectives on future samples of RPE stars and their dynamical analyses, and how new approaches to chemical-evolution modeling may help refine our understanding of nature and origin of the astrophysical $r$-process. 

\renewcommand{\arraystretch}{1.0}
\setlength{\tabcolsep}{0.5em}

\begin{deluxetable*}{lrrrllrrr}[t]
\tabletypesize{\footnotesize}
\tablecaption{Positions, Magnitudes, Distances, Radial Velocities, and Proper Motions for the Initial Sample\label{tab2_stub}}
\tablehead{
\colhead{Name} & \colhead{RA} & \colhead{DEC} & \colhead{$V$ mag} & \colhead{$d_{Gaia}$ (err)} & \colhead{$d_{\rm SH}$ (err)} & \colhead{Rad Vel} & \colhead{PM$_{\rm RA}$} & \colhead{PM$_{\rm DEC}$} \\
\colhead{} & \colhead{(J2000)} & \colhead{(J2000)} &  & \colhead{(kpc)} & \colhead{(kpc)} & \colhead{(km s$^{-1}$)} & \colhead{(mas yr$^{-1}$)} & \colhead{(mas yr$^{-1}$)}
}
\startdata
2MASS J00002259$-$1302275      & 00:00:22.60 & $-$13:02:27.5 & 12.88 & 5.71 (1.54)*    & 4.78 (1.47)**  & $-$104.4                      & $-$2.61    & $-$9.45    \\
2MASS J00002416$-$1107454      & 00:00:24.16 & $-$11:07:45.4 & 12.04 & 3.76 (0.68)     & 3.51 (0.79)*   & $-$110.4                      & $-$0.63    & $-$23.48   \\
SMSS J000113.96$-$363337.9     & 00:01:13.99 & $-$36:33:38.1 & 13.91 & 8.02 (2.24)*    & 7.28 (2.03)*   & 243.3 \textsuperscript {a}                 & 8.35     & $-$5.16    \\
HD 224930                    & 00:02:11.18 & $+$27:04:40.0 & 5.67  & 0.01 (0.00)     & 0.01 (0.00)    & $-$35.6                       & 723.11   & $-$933.75  \\
2MASS J00021668$-$2453494      & 00:02:16.70 & $-$24:53:49.5 & 13.49 & 4.86 (0.74)     & 5.11 (1.36)*   & 80.1                        & $-$3.61    & $-$8.09    \\
BPS CS 22957$-$0036            & 00:03:31.20 & $-$04:44:19.0 & 14.41 & 1.26 (0.09)     & 1.20 (0.15)    & $-$154.5 \textsuperscript {b}  & 38.64    & $-$22.27   \\
HD 20                        & 00:05:15.48 & $-$27:16:18.8 & 9.01  & 0.50 (0.01)     & 0.50 (0.02)    & $-$56.5                       & 132.43   & $-$39.92   \\
2MASS J00073817$-$0345509      & 00:07:38.17 & $-$03:45:50.9 & 11.65 & 2.86 (0.56)     & 2.72 (0.65)*   & $-$146.3                      & 2.30     & $-$16.04   \\
BPS CS 31070$-$0073            & 00:08:51.83 & $+$05:26:17.1 & 14.53 & $\dots$             & 1.44 (0.57)**  & $-$86.0 \textsuperscript {c}            & 19.87    & $-$12.43   \\
2MASS J00093394$-$1857008      & 00:09:33.94 & $-$18:57:00.8 & 11.18 & 1.26 (0.08)     & 1.29 (0.08)    & $-$71.4                       & 39.33    & $-$37.43  
\enddata

\tablecomments{Sources for (non-\textit{Gaia}) radial-velocity information:}

\tablecomments{\textsuperscript{a}~\citet{jacobson2015}, \textsuperscript{b}~\citet{roederer2014b}, \textsuperscript{c}~T.C. Beers, private comm., \textsuperscript{d}~\citet{hansen2015}, \textsuperscript{e}~\citet{beers1992}, \textsuperscript{f}~\citet{beers1995}, \textsuperscript{g}~\citet{barklem2005}, \textsuperscript{h}~\citet{ezzeddine2020}, \textsuperscript{i}~\citet{hansen2018}, \textsuperscript{j}~\citet{aoki2002}, \textsuperscript{k}~\citet{bonifacio2009}, \textsuperscript{l}~\citet{kunder2017}, \textsuperscript{m}~\citet{xing2019}, \textsuperscript{n}~\citet{carrera2013}, \textsuperscript{o}~\citet{cohen2013}, \textsuperscript{p}~\citet{holmbeck2020}, \textsuperscript{q}~\citet{aoki2005}, \textsuperscript{r}~\citet{frebel2006}, \textsuperscript{s}~\citet{sakari2018a}, \textsuperscript{t}~\citet{howes2015}, \textsuperscript{u}~\citet{howes2016}, \textsuperscript{v}~\citet{allende2000}, \textsuperscript{w}~\citet{aoki2010}}


\tablecomments{Source for (non-\textit{Gaia}) proper motions: \textsuperscript{1}~\citet{kharchenko2009}}

\tablecomments{Single (*) and double (**) marks indicate stars with relative distance errors in the range  $20\% < \epsilon \leq 30\%$, and $\epsilon > 30\%$, respectively.}

\tablecomments{This table is a stub; the full table is available in electronic form.}

\vspace{-20px}
\end{deluxetable*}

\section{Data Assembly, Distances, and Elemental Abundances}

We select 141 $r$-II, 332 $r$-I, and 46 limited-$r$ stars from a number of sources: the first four data releases from the RPA \citep{hansen2018,sakari2018a,ezzeddine2020,holmbeck2020}, the JINAbase literature compilation \citep{abohalima2018}, and a few additional stars from the recent literature that are not yet in the JINAbase compilation.  These are then cross-matched with \textit{Gaia} DR2 \citep{gaia2016,gaia2018a,gaia2018b,gaia2018c} in order to obtain parallaxes, proper motions, and radial velocities, where available. When this information is not available from \textit{Gaia}, we supplement it with  radial velocities from RAVE DR5, where available \citep{kunder2017}, as well as from other literature sources.  A minority of our stars ($\sim 3$\%) only had negative parallaxes from \textit{Gaia}; their distance estimates are obtained as described below. We further require that each of the stars have spectroscopic determinations of [Fe/H], [Ba/Fe], and [Eu/Fe], as these are required to make confident sub-class assignments. Stars with only reported upper limits on [Eu/Fe] are rejected, as are stars with [Fe/H] $ > -1.0$. Application of these cuts reduces our sample to 519 stars, which we refer to below as the Initial Sample. 

Distance estimates for our stars were obtained from inversion of the reported \textit{Gaia} DR2 parallax, with a positive correction of $0.054$ mas added to the parallax value, as recommended by \citet{everall2019} and \citet{schonrich2019}.  We also calculate distances using the Bayesian tool \texttt{StarHorse} \citep{queiroz2018,queiroz2020,anders2019} , one advantage of which is \texttt{StarHorse}'s ability to provide reasonable estimates of distances even for stars with negative (or missing) parallax values reported by \textit{Gaia} DR2. The \texttt{StarHorse} algorithm uses $T_\text{eff}$, $\log g$, relative chemical abundances, photometric magnitudes, and parallaxes, when available, as inputs, but it does not explicitly apply a correction to the \textit{Gaia} parallax values.

The resulting basic information for our program stars is listed in Table~\ref{tab2_stub}. Stars with relative errors  in their derived distances ($\epsilon = \Delta\,d/d$) in the range $20\% < \epsilon \leq 30\%$, and those with $\epsilon > 30$\%, are indicated in the table. We also list our adopted radial velocities, taken from \textit{Gaia} DR2 (with accuracy typically better than 1 km~s$^{-1}$), where available. When not, we adopt the radial velocities reported by high-resolution spectroscopic studies (with accuracy no worse than 1-2 km~s$^{-1}$ \citep{holmbeck2020}, or, in two cases, from medium-resolution studies (with accuracy no worse than 10 km~s$^{-1}$) for each star, as indicated in the table. Note that our preference for the choice of radial velocities from \textit{Gaia}, as opposed to other sources, is that we wished to preserve the homogeneity of our uncertainties, to the extent possible.   All but one of our stars have proper motions available from \textit{Gaia} DR2; in this one case we adopt a measurement from an alternative source, as indicated in the table.

\begin{figure}
    \includegraphics[width=\columnwidth]{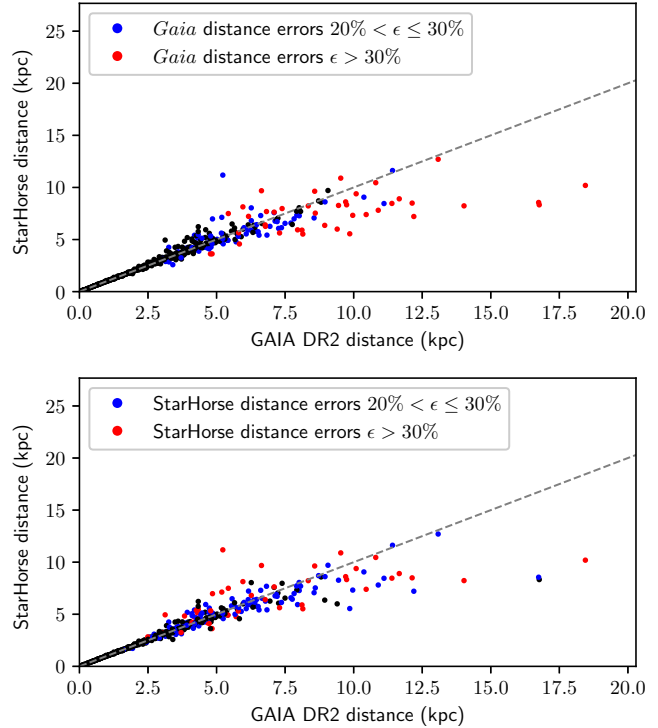}
	\caption{Comparison of \textit{Gaia} DR2 inverse-parallax distances and Bayesian \texttt{StarHorse} distances. The top panel marks stars with relative distance errors in the range 20\% $< \epsilon < 30$\%   with blue dots, and those with $\epsilon > 30$\% with red dots, respectively.  The bottom panel shows the same information,  but with the blue and red colored dots reflecting the stars with $20\% < \epsilon \leq 30\%$  and $\epsilon > 30$\% in the \texttt{StarHorse} distances, respectively. }
	\label{distance_comparison_paper}
\end{figure}

Figure~\ref{distance_comparison_paper} shows a comparison between the derived \textit{Gaia} DR2 and \texttt{StarHorse} distances. The top panel shows the stars with \textit{Gaia} relative distance errors in the range $20\% < \epsilon \leq 30\%$   with blue dots, and those with $\epsilon > 30$\% with red dots, respectively.  The bottom panel shows the same information, but with the blue and red dots reflecting the stars with $20\% < \epsilon \leq 30\%$  and $\epsilon > 30$\% in the \texttt{StarHorse} distances, respectively. From inspection, the two sets of distances are quite close to one another in the distance range $d < 5 $ kpc, and reasonably consistent for $d < 7.5$ kpc. If a star has an available distance estimate from \texttt{StarHorse} with $\epsilon \leq 30$\% (which applies to $\sim 95\%$ of the stars), we adopt it; otherwise we adopt the {parallax-corrected} \textit{Gaia} distance estimate, provided its $\epsilon \leq 30$\%. We remove all stars for which our best available distance estimate has $\epsilon > 30$\%. This results in a total of $466$ stars.  We carry the stars with relative distance errors $20\% < \epsilon \leq 30\%$ through our dynamical analysis below, but point out a number of stars ($113$) that may be suspect, due to the errors in their derived dynamical quantities arising from their uncertain distances.  We have gone to some length to preserve the stars with larger distance errors in our sample, in the expectation that improved distance estimates for most of them will come available in future \textit{Gaia} releases.

\renewcommand{\arraystretch}{1.0}
\setlength{\tabcolsep}{0.45em}

\begin{deluxetable*}{lllrrrrrrcl}[t]
\tabletypesize{\small}
\tablecaption{RPA Classes, Temperatures, and Elemental Abundances for the Initial Sample\label{tab3_stub}}
\tablehead{
 \colhead{Name} & \colhead{RPA Class} & \colhead{$T_{\rm eff}$} & \colhead{[Fe/H]} & \colhead{[C/Fe]} & \colhead{[C/Fe]$_c$} & \colhead{[Sr/Fe]} & \colhead{[Ba/Fe]} & \colhead{[Eu/Fe]} & \colhead{Reference} & \colhead{RPA} \\
 \colhead{} & \colhead{} & \colhead{(K)} & \colhead{} & \colhead{} & \colhead{} & \colhead{} & \colhead{} & \colhead{} & \colhead{} & \colhead{}
 }
\startdata
2MASS J00002259$-$1302275      & $r$-I           & 4576 & $-$2.90 & $-$0.65 & $-$0.65 & $-$1.20 & $-$0.38 & $+$0.58 & \textsuperscript {a}   & 1 \\
2MASS J00002416$-$1107454      & $r$-I           & 4693 & $-$2.43 & $-$0.27 & $+$0.30 & $-$0.04 & $-$0.24 & $+$0.51 & \textsuperscript {b}   & 1 \\
SMSS J000113.96$-$363337.9     & limited-$r$     & 4810 & $-$2.32 & $-$0.33 & $+$0.09 & $-$0.04 & $-$1.21 & $-$0.55 & \textsuperscript {c}   & 0 \\
HD 224930                    & $r$-I           & 5275 & $-$1.00 & $\dots$   & $\dots$   & $\dots$   & $-$0.19 & $+$0.34 & \textsuperscript {d}   & 0 \\
2MASS J00021668$-$2453494      & $r$-I           & 5020 & $-$1.81 & $-$0.88 & $-$0.57 & $+$0.59 & $+$0.10 & $+$0.52 & \textsuperscript {a}   & 1 \\
BPS CS 22957$-$0036            & $r$-I           & 5970 & $-$2.28 & $+$0.04 & $+$0.04 & $+$0.20 & $-$0.22 & $+$0.55 & \textsuperscript {e}  & 0 \\
HD 20                        & $r$-II          & 5445 & $-$1.58 & $-$0.34 & $-$0.32 & $+$0.13 & $+$0.32 & $+$0.80 & \textsuperscript {f}   & 0 \\
2MASS J00073817$-$0345509      & $r$-II          & 4663 & $-$2.09 & $-$0.32 & $+$0.17 & $+$0.41 & $+$0.11 & $+$0.73 & \textsuperscript {g}   & 1 \\
BPS CS 31070$-$0073            & $r$-II / CEMP-$r$ & 6190 & $-$2.55 & $+$1.34 & $+$1.34 & $-$0.04 & $+$2.42 & $+$2.83 & \textsuperscript {h}   & 0 \\
2MASS J00093394$-$1857008      & $r$-I           & 4815 & $-$1.88 & $-$0.17 & $+$0.06 & $+$0.13 & $+$0.23 & $+$0.46 & \textsuperscript {b}   & 1
\enddata
\tablecomments{The code in the column RPA indicates if the reported information is drawn from papers from the RPA (1) or other sources (0).}

\tablecomments{Sources for chemical-abundance information:}

\tablecomments{\textsuperscript{a}~\citet{hansen2018}, \textsuperscript{b}~\citet{holmbeck2020}, \textsuperscript{c}~\citet{jacobson2015}, \textsuperscript{d}~\citet{fulbright2000}, \textsuperscript{e}~\citet{roederer2014a}, \textsuperscript{f}~\citet{barklem2005}, \textsuperscript{g}~\citet{sakari2018a}, \textsuperscript{h}~\citet{allen2012}, \textsuperscript{i}~\citet{ezzeddine2020}, \textsuperscript{j}~\citet{hansen2015}, \textsuperscript{k}~\citet{christlieb2004}, \textsuperscript{l}~\citet{siqueira2014}, \textsuperscript{m}~\citet{Hansen2012}, \textsuperscript{n}~\citet{lai2008}, \textsuperscript{o}~\citet{rasmussen2020}, \textsuperscript{p}~\citet{preston2006}, \textsuperscript{q}~\citet{hollek2011}, \textsuperscript{r}~\citet{cayrel2004}, \textsuperscript{s}~\citet{cain2018}, \textsuperscript{t}~\citet{holmbeck2018}, \textsuperscript{u}~\citet{mardini2019}, \textsuperscript{v}~\citet{li2015}, \textsuperscript{w}~\citet{xing2019}, \textsuperscript{x}~\citet{valentini2019}, \textsuperscript{y}~\citet{hayek2009}, \textsuperscript{z}~\citet{ishigaki2013}, \textsuperscript{aa}~\citet{johnson2002}, \textsuperscript{ab}~\citet{cohen2013}, \textsuperscript{ac}~\citet{westin2000}, \textsuperscript{ad}~\citet{roederer2010}, \textsuperscript{ae}~\citet{aoki2005}, \textsuperscript{af}~\citet{hawkins2018}, \textsuperscript{ag}~\citet{burris2000}, \textsuperscript{ah}~\citet{honda2004}, \textsuperscript{ai}~\citet{cain2020}, \textsuperscript{aj}~\citet{cowan2002}, \textsuperscript{ak}~\citet{howes2015}, \textsuperscript{al}~\citet{johnson2013}, \textsuperscript{am}~\citet{placco2020}, \textsuperscript{an}~\citet{howes2016}, \textsuperscript{ao}~\citet{mcwilliam1995}, \textsuperscript{ap}~\citet{placco2017}, \textsuperscript{aq}~\citet{sneden2003}, \textsuperscript{ar}~\citet{mashonkina2014}, \textsuperscript{as}~\citet{ivans2006}, \textsuperscript{at}~\citet{ryan1996}, \textsuperscript{au}~\citet{roederer2018b}, \textsuperscript{av}~\citet{masseron2012}, \textsuperscript{aw}~\citet{aoki2010}}

\tablecomments{This table is a stub; the full table is available in electronic form.}

\vspace{-20px}
\end{deluxetable*}

Table~\ref{tab3_stub} lists the $r$-process-enhancement classes used by the RPA (as in  Table 1), the reported effective temperatures ($T_{\rm eff}$), and selected elemental abundances, along with the primary references used for their determination.  The last column provides an indicator, set to 1, if the star is included in the RPA data releases (or other RPA papers), or set to 0, if not. 

For the measured carbon abundances reported in Table~\ref{tab3_stub} ([C/Fe]), we apply the evolutionary correction developed by \citet{placco2014} in order to approximate the initial carbon abundance that occurs due to CN processing during the upper red giant-branch stage. The corrected abundance is referred to as [C/Fe]$_{c}$ in this and later tables.

\renewcommand{\arraystretch}{1.0}
\setlength{\tabcolsep}{0.65em}

\begin{deluxetable*}{lcrcccrrr}
\tabletypesize{\small}
\tablecaption{ Derived Dynamical Parameters for the RPE Sample\label{tab4_stub}}
\tablehead{
 \colhead{Name} & \colhead{Energy} & \colhead{Ecc} & \colhead{$J_r$} & \colhead{$J_\phi$}  & \colhead{$J_z$} & \colhead{$r_\text{peri}$} & \colhead{$r_\text{apo}$} & \colhead{$Z_\text{max}$} \\
 \colhead{} &  \colhead{$\times 10^5$} & \colhead{} & \colhead{$\times 10^3$} & \colhead{$\times 10^3$} & \colhead{$\times 10^3$} & \colhead{} & \colhead{} & \colhead{}\\
 \colhead{} & \colhead{(km$^2$ s$^{-2}$)} & \colhead{} & \colhead{(kpc km s$^{-1}$)} & \colhead{(kpc km s$^{-1}$)}  & \colhead{(kpc km s$^{-1}$)} & \colhead{(kpc)} & \colhead{(kpc)} & \colhead{(kpc)}
 }
 \startdata
2MASS J00002259$-$1302275*     & $-$1.50 & 0.79 & 0.74  & $+$0.57 & 0.34 & 1.68  & 14.46   & 9.65   \\
2MASS J00002416$-$1107454*     & $-$1.39 & 0.73 & 0.85  & $-$0.87 & 0.46 & 2.85  & 18.29   & 12.07  \\
HD 224930                    & $-$1.68 & 0.25 & 0.07  & $+$1.41 & 0.01 & 4.86  & 8.13    & 0.42   \\
2MASS J00021668$-$2453494*     & $-$1.42 & 0.40 & 0.25  & $+$0.92 & 1.07 & 6.18  & 14.51   & 12.61  \\
BPS CS 22957$-$0036            & $-$1.72 & 0.84 & 0.58  & $-$0.32 & 0.10 & 0.82  & 9.44    & 4.26   \\
HD 20                        & $-$1.57 & 0.98 & 1.15  & $+$0.06 & 0.01 & 0.15  & 13.44   & 5.68   \\
2MASS J00073817$-$0345509*     & $-$1.71 & 0.86 & 0.75  & $+$0.04 & 0.10 & 0.73  & 10.08   & 3.45   \\
2MASS J00093394$-$1857008      & $-$1.74 & 0.64 & 0.37  & $-$0.74 & 0.04 & 1.96  & 8.96    & 1.37   \\
2MASS J00101758$-$1735387      & $-$1.54 & 0.04 & 0.00  & $+$0.68 & 1.14 & 7.97  & 8.64    & 7.95   \\
2MASS J00154806$-$6253207      & $-$1.69 & 0.85 & 0.52  & $-$0.16 & 0.40 & 0.79  & 9.53    & 8.25   
\enddata 
\tablecomments{Single (*) mark indicates the {16 stars} 
with relative distance errors in the range  20\%$ < \epsilon < 30$\%. See 
text for details.}   

\tablecomments{This table is a stub; the full table is available in electronic form.}

\vspace{0px}
\end{deluxetable*}

For the remaining analysis we remove the limited-$r$ stars, which are relatively few in number ($46$), and may have nucleosynthetic origins that differ from the $r$-I and $r$-II stars \citep{frebel2018}. This rejection, along with the removal of stars having high relative distance errors described above, results in a final sample of 127 $r$-II and 319 $r$-I stars (for a total of 446 stars), which we refer to as the RPE Sample.

\begin{figure} 
    \includegraphics[width=\columnwidth]{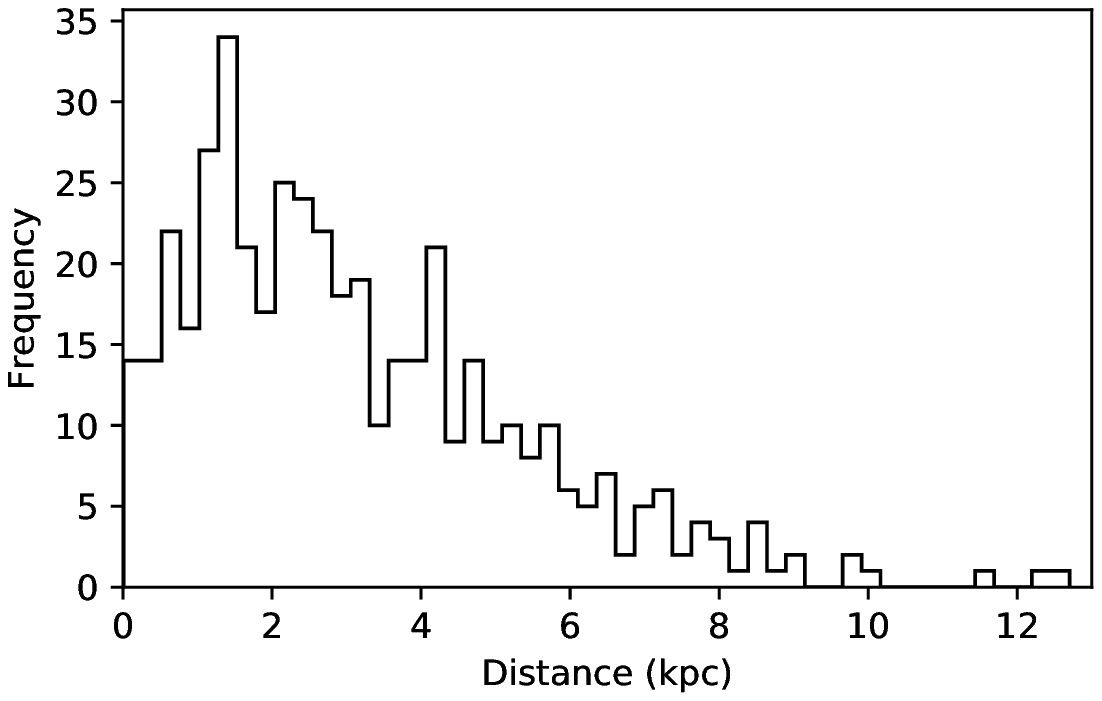}
	\caption{Adopted distances for the 446 stars in the RPE Sample, based on the \texttt{StarHorse} or \textit{Gaia} DR2} estimates. See text for details.
	\label{dist_paper}
\end{figure}

\begin{figure}[htpb]
    \includegraphics[width=\columnwidth]{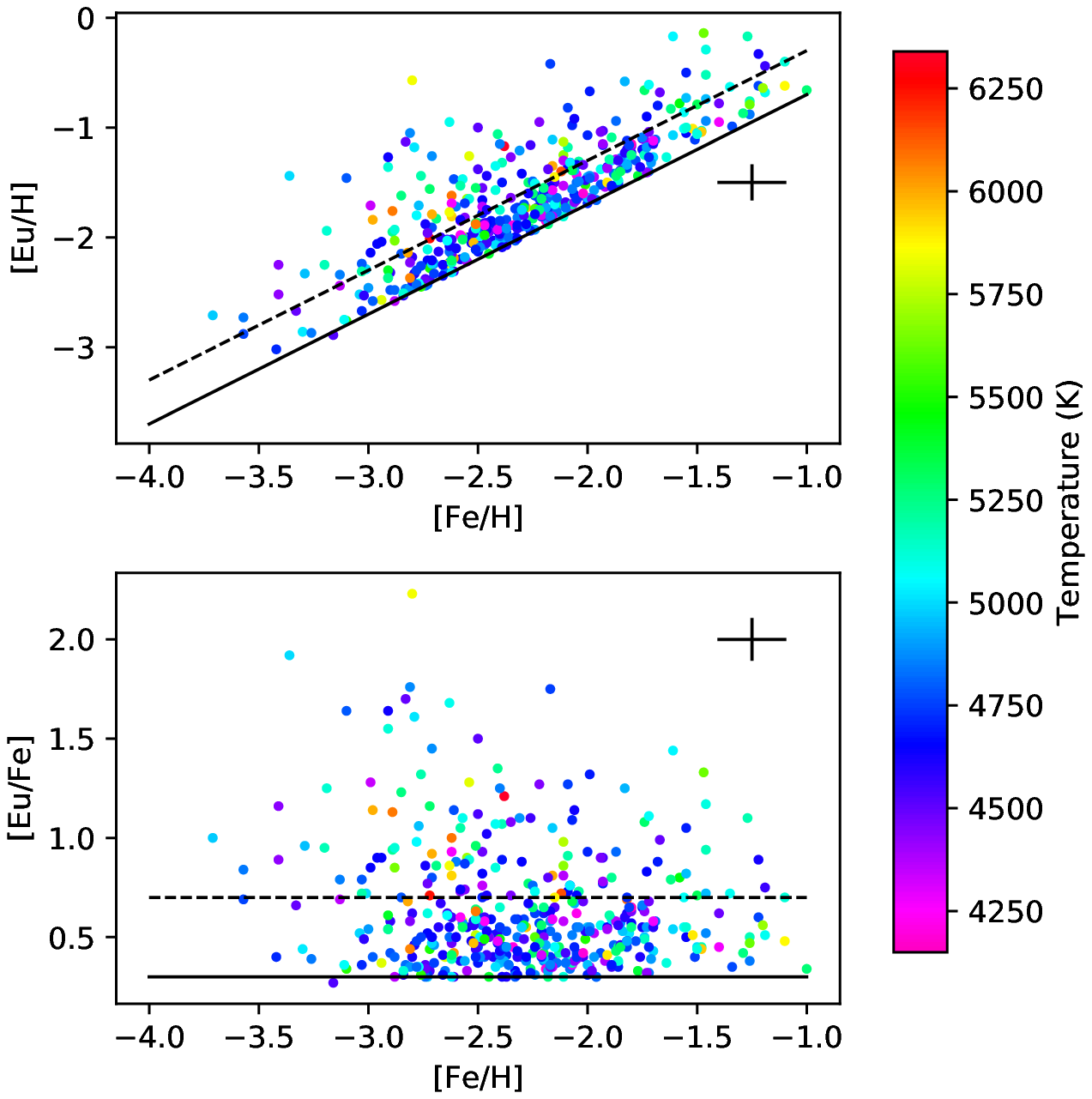}
	\caption{Europium abundances ([Eu/H], upper panel; [Eu/Fe], lower panel), as a function of metallicity ([Fe/H]), for the RPE Sample.  The solid black line denotes the [Eu/Fe] = +0.3 cutoff used by the RPA to include $r$-I stars; the dashed black line indicates the adopted division line for $r$-II stars, [Eu/Fe] = +0.7. Effective temperatures ($T_{\rm eff}$) are color-coded using the scale shown in the right margin. Representative error bars on the abundances are shown in each panel.}
	\label{Eu_FeH_paper}
\end{figure}

Figure \ref{dist_paper} shows the adopted distances for our RPE Sample stars.  More than $75\%$ of our stars are located within 5 kpc of the Sun; only four have distances more than 10 kpc. We recognize that estimates of the energies and other dynamical parameters based on poorly estimated distances can lead to significant errors. For this reason, \citet{roederer2018a} employed a conservative relative-parallax error cut of 12.5\%, reducing their original sample from 83 to 35 stars. As mentioned above, we have chosen to retain stars with relative distance errors $20\% < \epsilon \leq 30\%$, but removed stars with relative distance errors $\epsilon > 30$\%. 

Figure~\ref{Eu_FeH_paper} shows the distribution of [Eu/H] and [Eu/Fe] for the RPE Sample, as a function of [Fe/H], color-coded by $T_{\rm eff}$. The metallicity range of our stars extends over $-3.7 < $ [Fe/H] $< -1.0$. 
Most stars have $T_{\rm eff}$ in the range [4400, 5200]~K, which is essentially set by our demand for a measured [Eu/Fe]; there are a few $r$-II stars in the RPE sample with temperatures as high as $T_{\rm eff} \sim 6650$~K. The coolest stars included in the RPE sample have $T_{\rm eff}\sim 4150$~K.  Note that, because of the inclusion of $r$-I stars, the range in [Eu/Fe] we consider in the RPE Sample, $+0.30 < $ [Eu/Fe] $ \leq +2.28$, is substantially larger than that considered by \citet{roederer2018a}.

\section{Estimated Dynamical Parameters}

We derive orbital energies and other integrals of motion (actions) for the RPE sample, along with other dynamical parameters, using the \texttt{AGAMA} package \citep{vasiliev2018} and the MW2017 potential \citep{mcmillan2017}\footnote{We used the axisymmetric potential, not accounting for the central rotating bar.}.  These results are listed in Table~\ref{tab4_stub}. We note that there exist a number of sources of errors, both statistical and systematic, that contribute to the overall uncertainty in the estimation of these parameters. This list includes, but is not limited to, the measurement errors in the distance estimates to each star, errors in the reported proper motions, errors in the radial velocities, and covariances between these quantities, as well as systematics arising from the choice of Galactic potential, and other unrecognized systematic errors. Errors in the distance estimates are very likely, however, to be the dominant contributor.  These are listed, star-by-star, in Table 2.

\begin{figure}[t]
    \includegraphics[width=\columnwidth]{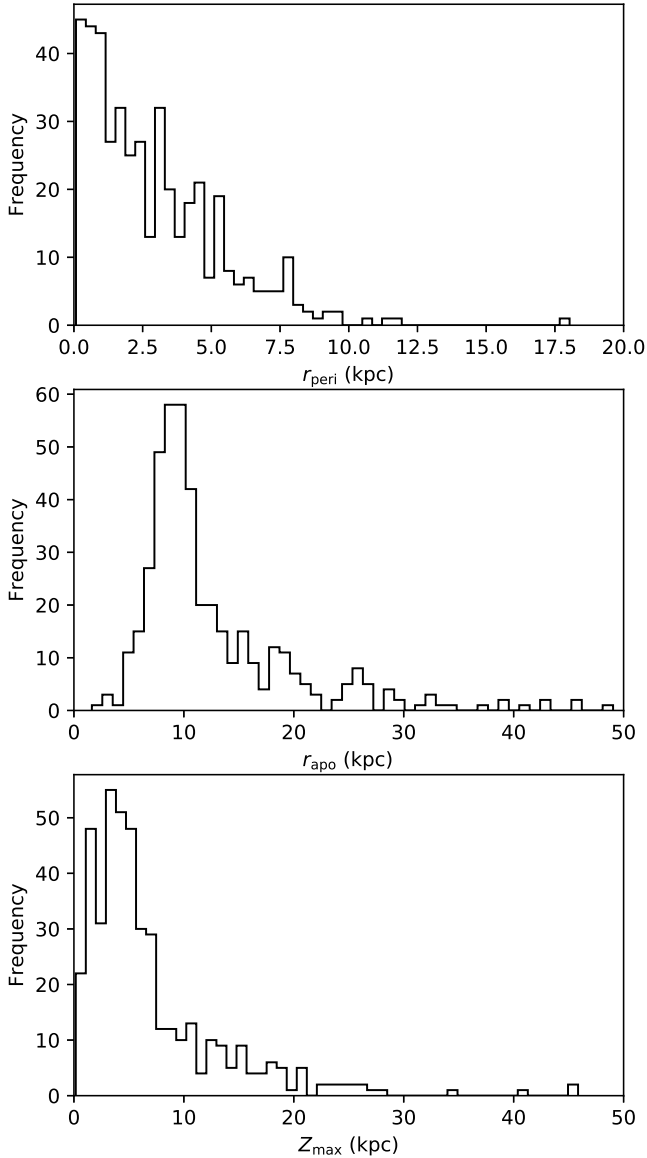}
	\caption{Distribution of pericentric distances, $r_{\rm peri}$ (top panel), apocentric distances, $r_{\rm apo}$ (middle panel), and maximum distance from the Galactic plane achieved during the stars' orbits, $Z_{\rm max}$ (bottom panel), for stars in the RPE Sample. Note that 10 stars with $r_{\rm apo} > 50$ kpc and 7 stars with $Z_{\rm max} > 50$ kpc are not shown.}
	\label{distances_paper}
\end{figure}

Figure~\ref{distances_paper} shows histograms of the pericentric distances, $r_{\rm peri}$, apocentric distances, $r_{\rm apo}$, and the maximum distance from the Galactic plane achieved during the stars' orbits, $Z_{\rm max}$. \texttt{AGAMA} calculates the stellar orbits by integrating the motions of each star over a period of 3 Gyr (from its present position), from which these values are obtained.  The orbital eccentricity is obtained from $e = (r_{\rm apo} - r_{\rm peri})/(r_{\rm apo} + r_{\rm peri})$.

It is noteworthy that all of the pericentric distances of the RPE stars are within 20 kpc, as was also pointed out by \citet{roederer2018a} for their much smaller sample.  This has clear implications for the likely birth environments of the RPE stars, as discussed in more detail below.

\begin{figure*}
    \includegraphics[width=\textwidth]{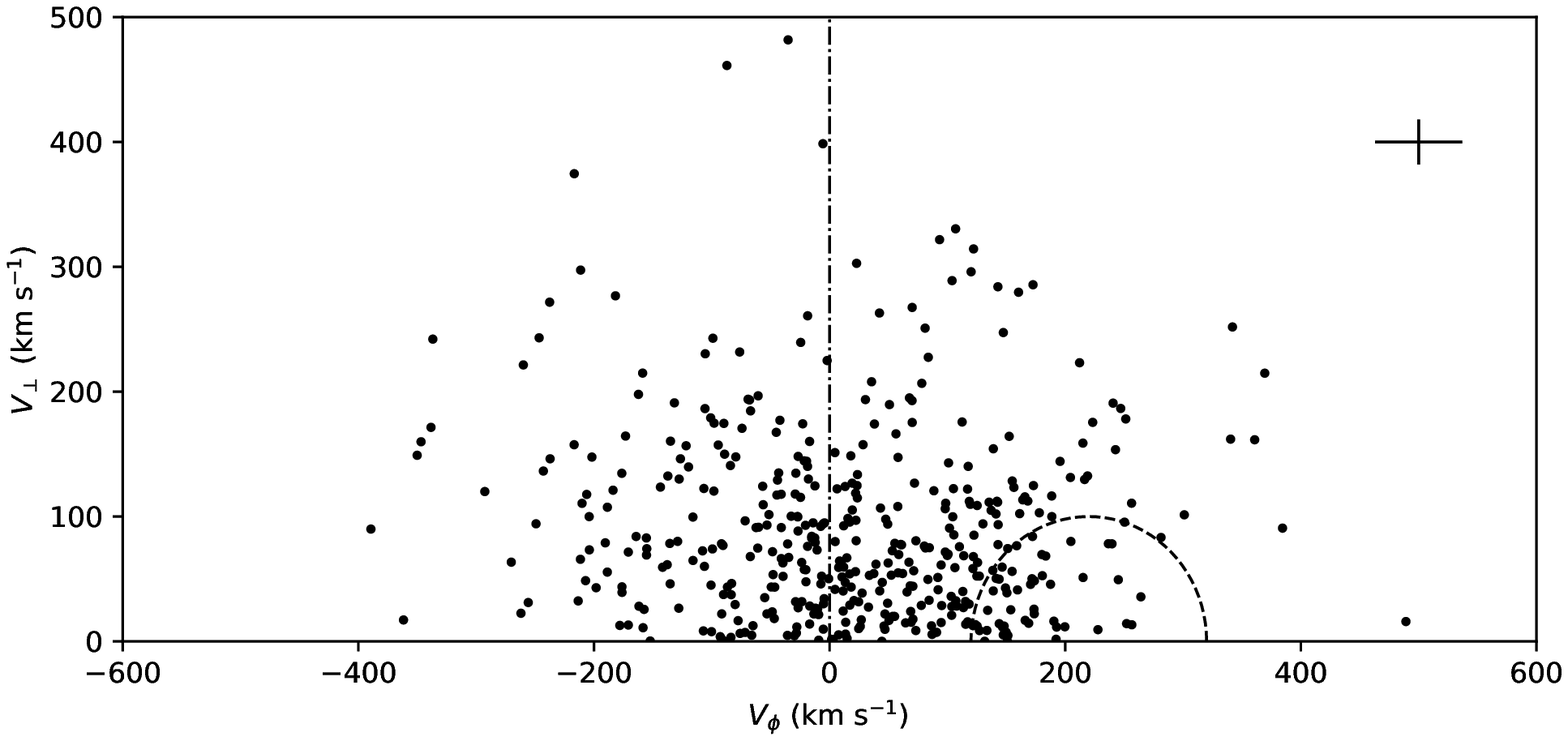}

\caption{Toomre diagram for the RPE Sample. The dashed semi-circle denotes the radius of 100 km s$^{-1}$ from the approximate location of the Local Standard of Rest, $V_\phi\approx 220$ km s$^{-1}$. The vertical dash-dotted line separates stars with prograde ($V_\phi>0$) orbits from those with retrograde ($V_\phi<0$) orbits. A representative error bar is shown in the upper right of this figure, multiplied by a factor of 5 for better visibility.} 
	\label{toomre_paper}
\end{figure*}

Figure~\ref{toomre_paper} is the Toomre diagram for the stars in the RPE Sample.  From inspection, all but 51 of the stars are clearly kinematically associated with the halo system, and lie outside the  semi-circular region indicating a 100 km~s$^{-1}$ velocity difference from the Local Standard of Rest, within which stars are potentially members of the disk system (all but 7 of these potential disk-system stars have $Z_{\max} < 4$ kpc).  Considering only the clear halo-system stars, those on prograde  orbits (201; $51\%$) are roughly equal in numbers to those on retrograde orbits (194; $49\%$).

The energies and the other integrals of motion (actions) are defined \citep{binney2012}  as:
\begin{itemize}
    \item  $E$, the specific orbital energy, defined as the full orbital energy of the star divided by its mass.
    \item $J_r$, the orbital action due to oscillations along the orbital radius, is defined to be non-negative.  For a given $E$, $J_r$ = 0 for circular orbits and is large for eccentric orbits. 
    \item $J_{\phi}$, the action due to orbital motions, is taken to be the reverse projection of the angular momentum on the axis perpendicular to the Galactic plane, $-L_z$. In our adopted coordinate system, positive and negative values of $J_{\phi}$ (and $V_{\phi}$) correspond to prograde and retrograde motions about the Milky Way's center, respectively.
    \item $J_z$, the action due to oscillations along the axis perpendicular to the Galactic plane, is defined to be non-negative.
    $J_z$ = 0 for planar orbits, and is large for orbits with large $Z_{\rm max}$. 
\end{itemize}

\begin{figure*}
	\includegraphics[width=\textwidth]{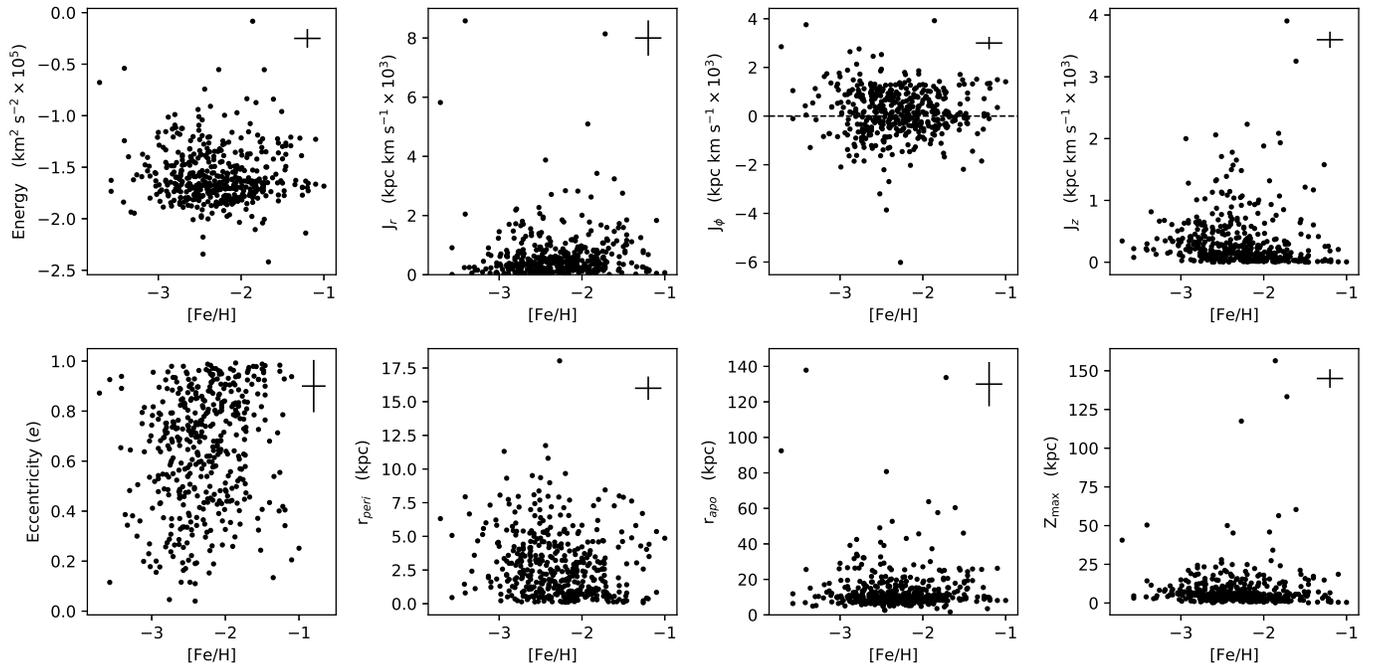}
	\caption{Orbital energies, actions, and other dynamical parameters for the RPE Sample, derived using \texttt{AGAMA} with the MW2017 potential, as a function of [Fe/H].  The horizontal dashed line in the $J_{\phi}$ vs. [Fe/H] plot divides stars with prograde from retrograde orbits. Horizontal error bars in the upper-right corners show characteristic uncertainties in the [Fe/H] determination, and the vertical error bars show mean uncertainties (multiplied by a factor of 5 for better visibility) of the involved quantities.} 
	\label{kinematics_paper}
\end{figure*}

Figure~\ref{kinematics_paper} shows the distributions of the derived energies and other integrals of motions (actions) for the RPE Sample, along with several additional dynamical parameters, as a function of [Fe/H]. From inspection of this figure, even though the RPE Sample comprises relatively bright stars (which was preferable in order to obtain high-resolution spectroscopic follow-up), the derived dynamical parameters cover a wide range of values, and also span a large range of [Fe/H].  If this were not the case, it would be unsurprising to find clusters in the dynamical parameters and low dispersions in [Fe/H], due to random selection alone.   

We note that, for the actual clustering procedure described in the next section, we remove four additional stars (BPS CS 29526-110, BPS BS 15621-047, BPS BS 16929-005, and BPS BS 16033-081) from the sample due to their abundances strongly varying across different studies, putting their classification shown in Table \ref{tab3_stub} into question.

\section{Clustering Procedure}

A number of clustering approaches have been explored in the past for considering the dynamical groupings of halo stars.  For example, \citet{helmi2000} ran theoretical simulations, and found that a Friends of Friends (FoF) approach, clustering over $E$, $L_z$, and total orbital momentum, $L$, can recover over 50\% of all galactic accretion events, indicating low variability over time of these integrals of motion. Other observational works employed a number of alternatives. These include: (1) density analysis and subsequent application of a so-called watershed algorithm \citep{roerdink2000} in the space of $E$ and $J_{\phi}$ \citep{helmi2017}, (2) visual density-map analysis in the space of $E$, $J_{\phi}$, and $L_{\perp}$ (the component of $L$ perpendicular to $L_z$) \citep{koppelman2018}, (3) \texttt{HDBSCAN} \citep{mcinnes2017,malzer2019}, used to explore clustering applied to the space of $E$, $J_{\phi}$, $e$, and [Fe/H] \citep{koppelman2019}, and (4) the four approaches explored by the \citet{roederer2018a} analysis of RPE stars -- K-means clustering (\citealt{lloyd1982}; \citealt{arthur2007}), agglomerative clustering (\citealt{ward1963}), affinity propagation clustering (\citealt{frey2007}), and mean-shift clustering (\citealt{comaniciu2002}), all applied to the space of $E$ and the orbital actions $J_r$, $J_{\phi}$, and $J_z$. Most recently, \citet{yuan2020} applied the neural-network based method \texttt{STARGO} \citep{yuan2018} to the $E$, $L$, $\theta$, and $\phi$ space, where the latter two angular parameters characterize the directions of the orbital poles. 

Each of the above clustering approaches have advantages and disadvantages, and different algorithms may be useful for different applications, and for data sets of varying size and possible cluster shapes. For example, \texttt{STARGO} appears to be particularly suitable for large ($N > 1000$) data sets, and in particular, provides an objective means for assessing the confidence (and possible contamination) of the various clusters it identifies. The methods employed by \citet{roederer2018a} were chosen because they were thought to be best for smaller ($N < 100$) data sets. Whatever approach is taken, there is always some subjective set of choices that must be made by the user, which can lead to complexity in the interpretation of the final results.

After conducting experiments with a number of the above techniques, we initially decided, in the interest of simplicity and reproducibility, as well as our concern over the limited size of the RPE Sample, to employ a FoF\footnote{The particular implementation used was the \textit{pyfof} Python library (\url{https://github.com/simongibbons/pyfof}).} algorithm in the space of $E$, $J_\phi$, and $e$. The FoF results depend on the choice of a single parameter, $l$, the linking length, which allowed us to explore how the elemental-abundance dispersions of the resulting dynamical clusters depended on this parameter. The results clearly indicated a statistically significant reduction of abundance dispersions in the identified dynamical clusters compared to clusters formed from random selections.  We were somewhat concerned (as was an anonymous referee), however, that our optimal choice of $l$ led to several very large dynamical clusters ``tied together" by single stars in low-density regions of the dynamical space.

We thus chose to implement a more powerful clustering algorithm, \texttt{HDBSCAN}, which has recently been employed to find DTGs of VMP stars from the HK \citep{beers1985,beers1992} and Hamburg/ESO \citep{christlieb2008} surveys by \citet{limberg2020}. Here, we combine the DTG-search methodology presented by these authors with our statistical tools (described below) to evaluate the chemical nature of the identified CDTGs. For application of \texttt{HDBSCAN}, we employed the same input dynamical parameters as used by \citet{roederer2018a} and \citet{limberg2020}: energy, $E$, and the three actions, $J_r$, $J_\phi$ and $J_z$. Note that we specifically {\it did not} choose to include [Fe/H], or any other chemical information, among the input parameters for this exercise, as we wished to quantitatively test the similarities of the elemental abundances for the members of a given cluster. This allows us to evaluate the hypothesis that the cluster members share a common chemical-evolution history.

The result of this procedure is the identification of 30 CDTGs, ranging in size from 3 to 12 stars, with 21 of the CDTGs containing between 3 and 5 stars each, and 9 containing between 6 and 12 stars each. The RPE stars in these CDTGs are listed in Table~\ref{tab5_stub}. In this table, the identified CDTGs are listed in order of declining confidence level (CL, described below).  Within each CDTG, the member stars are listed in order of increasing [Eu/Fe].  We discuss these CDTGs in more detail below. 

For convenience (e.g., for matching the CDTGs to other known substructures or dynamical groups), and for future reference, Table~\ref{tab6} lists the average energy, actions, eccentricity, and orbital rotation, $V_{\phi}$, along with the means and dispersions of the elemental abundances considered here.

\begin{figure*}[htp]
	\includegraphics[width=\textwidth]{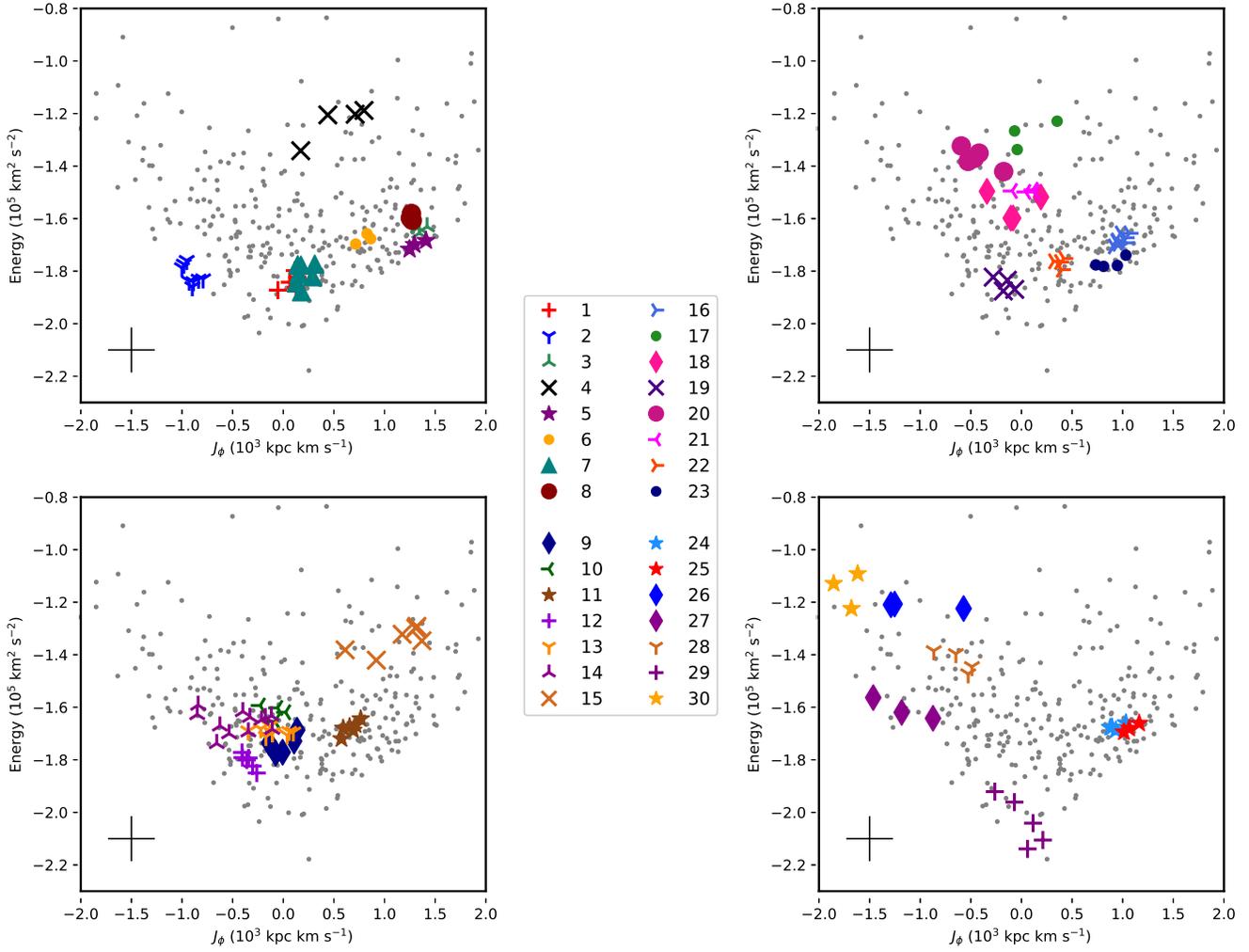}
	\caption{Energy vs. $J_\phi$ plots for the CDTGs in the RPE Sample. Different symbols and colors correspond to stars in different CDTGs, as shown in the legend.  Small gray dots are stars from the RPE Sample that have not been assigned membership to a CDTG.  The CDTGs are split into four separate plots for clarity. The error bars in the bottom-left corners show mean uncertainties (multiplied by a factor of 5 for better visibility) of the involved quantities.}
	\label{E_Jphi_paper}
\end{figure*}

Figure~\ref{E_Jphi_paper} shows plots of $E$ vs. $J_{\phi}$ for the CDTGs and non-clustered stars in the RPE Sample, as such plots have been used by numerous previous studies of DTGs, and are useful for comparisons with those works, as well as with future studies.  Although the plots are split into four individual panels, some apparent overlap in this dynamical space is unavoidable.  These are relieved for most cases in the projected action-space figures described below.

\begin{figure*}[htpb]
	\includegraphics[width=\textwidth]{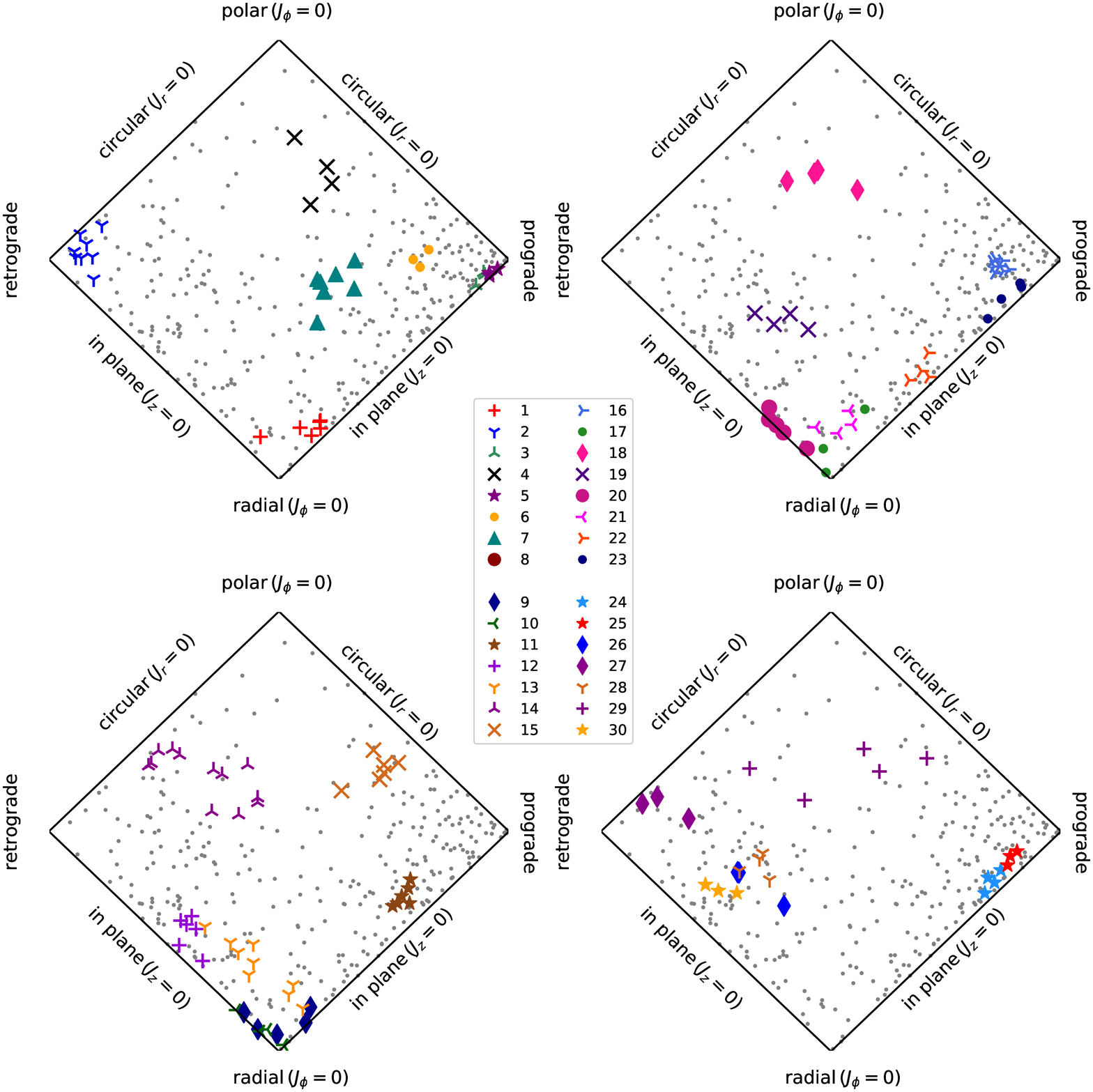}
	\caption{Projected action-space plots for the resulting clusters of stars from the RPE Sample. The x-axis is ($J_{\phi} / J_\text{tot}$),
		and the y-axis is $(J_z -J_r)/J_\text{tot}$, where $J_\text{tot} =  J_r + |J_{\phi}| + J_z$. Different symbols and colors correspond to stars in different CDTGs, as shown in the legend.   Small gray dots are stars from the RPE Sample that have not been assigned membership to a CDTG. The same markers are used for the same groups as in  Figures~\ref{E_Jphi_paper}. The CDTGs are split into four separate plots for clarity.}
	\label{rhombus_paper}
\end{figure*}

Figure~\ref{rhombus_paper} shows projected action-space maps for the CDTGs and the non-clustered stars in the RPE Sample.
From inspection of this figure, it is clear that the RPE CDTGs exhibit a variety of orbital behaviors, and variations in their degree of clustering (``tight" vs. ``relaxed").  Note that stars with prograde and retrograde orbits, and stars with radial and polar orbits, always appear together in the same CDTGs, with the exception of CDTG-29, which exhibits a noticeable spread in its orbital characteristics, from mildly retrograde to mildly prograde orbits.

For each of the obtained CDTGs, we also investigate their significance against the uncertainties of the dynamical parameters of their respective member stars. Mean errors for energies and actions are $1.3\%$ for $E$, $8.9\%$ for $J_r$, $2.3\%$ for $J_\phi$, and $7.0\%$ for $J_z$. We assume normal distributions for these quantities with its nominal value being the mean and its error the standard deviation of the distribution. We first produce $1,000$ sets of the 4-D energy-action vector of each star with a Monte Carlo procedure. Then, we re-evaluate the cluster assignment of these generated sets, and calculate the fraction of instances that each given star was attributed to the same CDTG. We take these fractions as representative of the membership probability of each star belonging to their assigned CDTG. Finally, for each CDTG, we average the resulting probabilities over all its member stars, and define this as the confidence level (CL) of the CDTG. The complete methodology for the \texttt{HDBSCAN} algorithm and CL estimation are described in \citet{limberg2020}.

The CL values are indicated for each CDTG in Table~\ref{tab5_stub}, where CDTGs are ordered by these values. Most of the CDTGs (23 out of 30) have CL values above $75\%$, and only 4 of them have CL values below $50\%$. This indicates the relative stability of our CDTGs to errors in the dynamical parameters, although improved distance estimates will certainly improve the accuracy of our results.

\section{The Chemical Nature of the Identified CDTGs}

We now explore the chemical properties of the identified CDTGs listed in Table~\ref{tab5_stub}, based on the limited set of elemental abundances provided for each star in the RPE Sample listed in Table~\ref{tab3_stub}. Our search is for evidence whether or not stars in the individual CDTGs exhibit elemental abundances that are more similar to one another than would have been obtained from considering the chemical behavior of similar-sized groups of stars chosen from the RPE Sample at random (that is, without considering their dynamical behavior). For this exercise, we obtain large numbers of mock groupings ($10^{6}$ per abundance per cluster size) of $N$ stars (matching the numbers of stars in the full set of CDTGs), randomly selected from the RPE Sample (where $3 \leq N \leq 12$), and derive the means and sample standard deviations of the elements considered (When the number of available measured elemental abundances is $4$ or more, biweight estimates (see, e.g., \citet{beers1990}) of these quantities are reported, in order to decrease the influence of potential outliers). Since the clustering was based entirely on dynamical parameters, and did not involve the stellar elemental abundances, we can quantify the degree to which the elemental abundances among members of each CDTG are more similar to one another than expected by chance by comparing to the sample standard deviations of the mock groups, as described below.  Note that, although we report values of the original measured carbonicity, [C/Fe], for all of stars in the CDTGs listed in Table~\ref{tab5_stub}, for the statistical analysis reported below we used the evolutionary-stage corrected value of the carbonicity, [C/Fe]$_c$, calculated following \citet{placco2014}.



Figure~\ref{dispersions_paper_cumuls} shows the cumulative distribution functions (CDFs) of the sample standard deviations (which we refer to as dispersions below) in these abundances for randomly drawn groups of $3$, $4$--$5$ and $6$--$12$ stars, respectively, from the 442 stars analyzed in the RPE Sample (Recall that four stars were dropped from consideration due to doubts about the reliability of their abundance determinations; see Section 3), indicated by the curved dashed lines in each panel (in the last two rows the boundaries of the region occupied by all the CDFs are shown). Each panel also indicates the numbers of stars in the CDTGs that are considered for each element. The horizontal dashed lines indicate the CDF values of $0.50$, $0.33$, and $0.25$ (corresponding to one-half, one-third, and one-quarter of the full CDF), respectively. We can then ask, element-by-element, whether or not the observed dispersions for stars in the CDTGs are distributed throughout the CDFs obtained for the full RPE Sample with these measurements (which varies slightly depending on the element under consideration) in a manner expected from chance, as described below.

For the elemental-abundance dispersions selected at random for CDTGs of a given size, the probability of the number of clusters lying below a given CDF value (from 0 to 1) is described by the binomial distribution. For $N$ total random-abundance dispersions, the probability that exactly $n$ of their CDF values lie below the value $v$ is:

\begin{equation}\label{eq1}
p_v(k=n,~N) = C_N^n v^n (1-v)^{N-n},
\end{equation}
\noindent where 
\begin{equation}\label{eq2}
C_N^n=\frac{N!}{(N-n)!n!}.
\end{equation}

The cumulative probability that $n$ or more CDF values lie below the value $v$ is found by summation of these probabilities over all values $k=n,n+1,\dots,N$:

\begin{equation}\label{eq3}
p_v(k\geq n,~N)=\sum\limits_{k=n}^N C_N^n v^n (1-v)^{N-n}.
\end{equation}

We refer to these probabilities as the Individual Elemental-Abundance Dispersion (IEAD) probabilities -- these are just the binomial probabilities for each element for the levels $v$ =  0.50, 0.33, and 0.25, respectively.  If the resulting IEAD probabilities for our observed numbers of clusters lying below certain values, $v$, are significantly below $50\%$, this is a strong indication that the abundance dispersions for a given element in our CDTGs do not result from chance.


Table \ref{tab7} summarizes the CDF statistics for all of the relevant abundances: [Fe/H], [C/Fe]$_c$, [Sr/Fe], [Ba/Fe], and [Eu/Fe]. The first column lists the abundance under consideration, the second displays the number of CDTGs that have a measured dispersion of this abundance (the requirement for which is the presence of at least three stars in a given CDTG with available measurements), the third shows the numbers of CDTGs with abundance-dispersion values in the CDF of dispersions obtained from the full RPE Sample below $v$ = 0.50, 0.33, and 0.25, respectively, and the last columns list the various probabilities of obtaining these (or larger) numbers of such CDTGs from random selection, as described below. 

\subsection{Individual Elemental-Abundance Behavior}

Here we consider the behavior of the individual elements within the CDTGs, on an element-by-element basis.  

\begin{figure*}[htpb]
	\includegraphics[width=\textwidth]{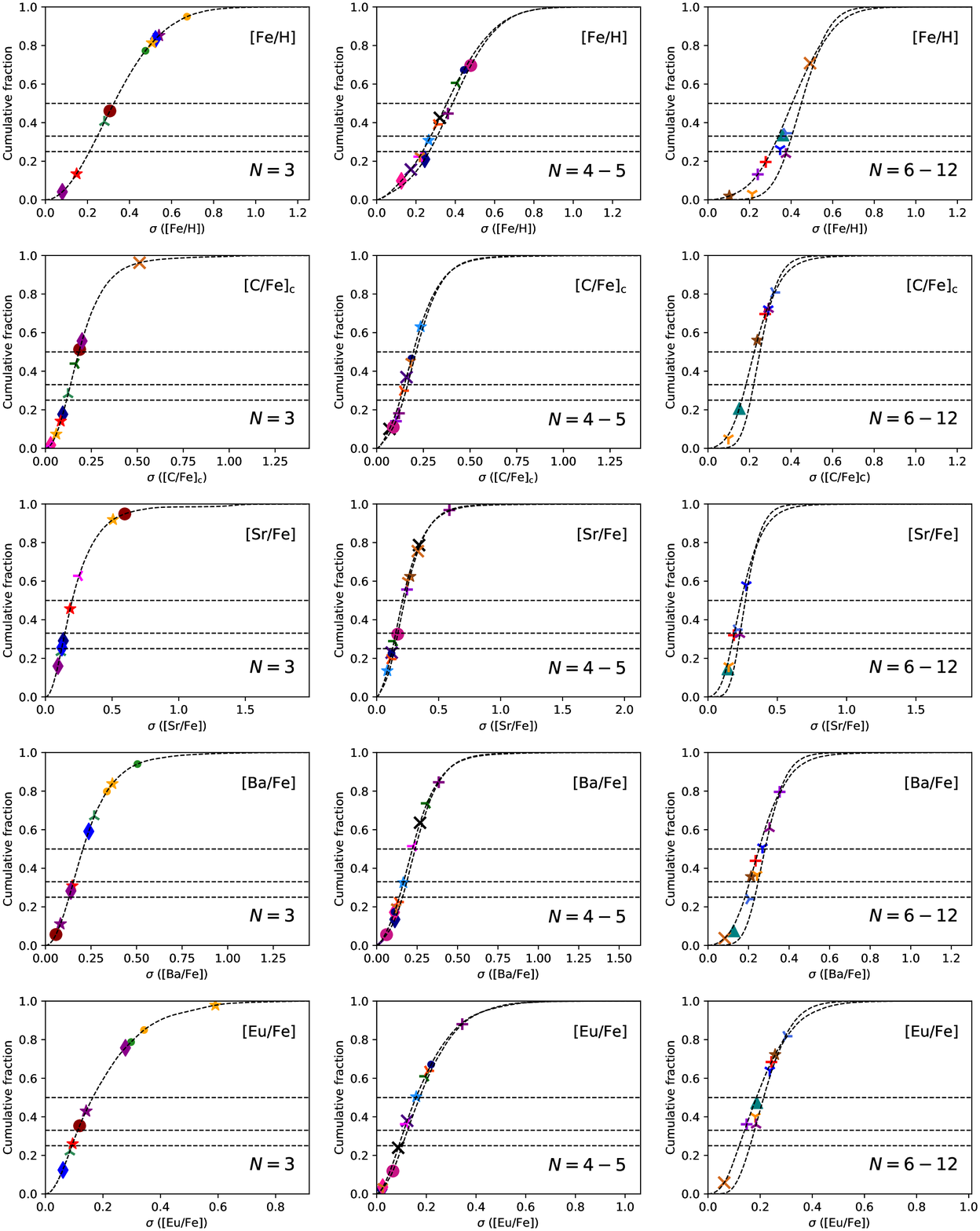}
	\caption{Dispersions in the elemental abundances within each CDTG, shown  on the CDFs of random groupings of all stars (with the same numbers of stars as the identified CDTGs) in the full RPE Sample. The left column includes CDTGs with 3 measured abundances. The next two  columns correspond to CDTGs with $4$--$5$ and $6$--$12$ measured abundances, respectively; all CDFs in these rows fall between the two curved dashed lines. The horizontal dashed lines show the $0.5$, $0.33$, and $0.25$ cumulative fraction values.}
	\label{dispersions_paper_cumuls}
\end{figure*}

{\bf [Fe/H] --}  The first row of Table~\ref{tab7} and the first row of panels in Figure~\ref{dispersions_paper_cumuls} summarize the dispersions of [Fe/H] within the CDTGs.  The dispersions are notably lower, compared to randomly drawn groups of stars from the CDFs of the full RPE Sample. All three IEAD (binomial) probabilities are below $10\%$, as 21 of the 30 identified CDTGs have CDF values below the 0.50 level, 14 are lower than the 0.33 level, and 12 of them are below the 0.25 level.

We can directly compare the more selective and smaller \citet{roederer2018a} sample and our more inclusive and larger sample using the above metric.  In the smaller sample, the binomial probability of finding 6 of 6 (5 of 6) groups with $N_{\rm star} \geq 3$ with [Fe/H] dispersions $<$ 0.50 (0.25) in the CDF is 1.6\% (0.4\%) by chance.  Only for the smallest [Fe/H] dispersions ($<$ 0.25 in the CDF) do those probabilities differ meaningfully from the ones found here with the expanded sample.  

{\bf [C/Fe]$_c$ --} The second row of Table~\ref{tab7} and the second row of panels in Figure~\ref{dispersions_paper_cumuls} summarize the dispersions of [C/Fe]$_c$ within the CDTGs. The dispersions for most  of the CDTGs (17/26) are lower than the 0.50 level for randomly drawn groups of stars from the CDFs of the full RPE Sample, resulting in a  binomial probability of 11.5\%.  Half of them are below the 0.33 level, and a bit under half are below the 0.25 level, resulting in small binomial probabilities of 8.4\% and 5.4\%. 

{\bf [Sr/Fe] --} The third row of Table~\ref{tab7} and the third row of panels in Figure~\ref{dispersions_paper_cumuls} summarize the dispersions of [Sr/Fe] within the CDTGs.
The dispersions for most of the CDTGs with measured [Sr/Fe] dispersions (16/26) are lower than the 0.50 level, compared to randomly drawn groups of stars from the CDFs of the full RPE Sample, resulting in a  binomial probability of 16.3\%. There are 15 with dispersions lower than the 0.33 level, resulting in a tiny binomial probability of 0.8\%, and 9 of them are below 0.25, resulting in a  binomial probability of 18.0\%.

{\bf [Ba/Fe] --} The fourth row of Table~\ref{tab7} and the fourth row of panels in Figure~\ref{dispersions_paper_cumuls} summarize the dispersions of [Ba/Fe] within the CDTGs.
The dispersions for 60\%  of the CDTGs (18/30) are lower than the 0.5 level for randomly drawn groups of stars from the CDFs of the full RPE Sample, resulting in a binomial probability of $18.1\%$.  Half of the CDTGs (15/30) are lower than the 0.33 level, resulting in a binomial probability of 4.0\%, and 12 of them are also below the 0.25 level, with a corresponding binomial probability of 5.1\%.
 
{\bf [Eu/Fe] --} The fifth row of Table~\ref{tab7} and the fifth row of panels in Figure~\ref{dispersions_paper_cumuls} summarize the dispersions of [Eu/Fe] within the CDTGs.
The dispersions for the majority  of the CDTGs (17/30) are lower than the 0.50 level for randomly drawn groups of stars from the CDFs of the full RPE Sample, resulting in a  binomial probability of only $29.2\%$.  Less than a third of the CDTGs (9/30) are lower than the 0.33 level, resulting in a statistically insignificant binomial probability of $70.1\%$. Almost a third (8/18) of them are below the 0.25 level, for a binomial probability of $48.6\%$.

\subsection{Full Elemental-Abundance Behavior}

We can make an estimate of the probability that our full set of elemental-abundance dispersions could have been arrived at through random draws from the CDFs for stars in the RPE Sample. To accomplish this, we take the five IEAD (binomial) probabilities ($p_1$, $p_2$, $p_3$, $p_4$, $p_5$), for the full set of five elements under consideration, listed in the fourth column of Table~\ref{tab7}. The first entry in that column is the $0.50$ CDF value, the second is the $0.33$ CDF value, and the third entry is the $0.25$ CDF value. 

As these probabilities correspond to different elements, they are expected to be independent of one another. We can thus characterize the statistical significance of obtaining them by estimating the probability that a set of these or lower probabilities (meaning a set of probabilities that, when ordered from the lowest to the largest, are each smaller than the corresponding probability from the set we have obtained) is generated by random draws of five probabilities from $0\%$ to $100\%$. This probability can be calculated in the same fashion as the GEAD probabilities in the next section (Eqn. \ref{eq7}): we simply replace $N$ with 5, $N_1$, $N_2$, $N_3$ with 1, 2, 3, 4, 5, and 0.50, 0.33, and 0.25 with our set of five probabilities, $p_1$, $p_2$, $p_3$, $p_4$, $p_5$, ordered from the smallest to the largest, to obtain the following formula for the Full Elemental-Abundance Distribution (FEAD) probability:

\begin{equation}\label{eq4}
\begin{split}
p_\text{FEAD} &= p(k_1\geq 1,~ k_2\geq 2,~ k_3\geq 3,~ k_4\geq 4,~ k_5\geq 5)\\
& = \sum\limits_{\mathclap{\substack{n_1,n_2,n_3,n_4,n_5\in\mathbb{Z},~~~~~~~~~\\n_1\geq 1,\\n_2\geq 2,\\n_3\geq 3,\\n_4\geq 4,\\n_5= 5,\\0\leq n_1\leq n_2\leq n_3\leq n_4\leq n_5= 5~~~~~~~~~~~}}}p(k_1,k_2,k_3,k_4,k_5) = (n_1,n_2,n_3,n_4,n_5),
\end{split}
\end{equation}

\noindent where:

\begin{equation}\label{eq5}
\begin{split}
p(&k_1=n_1,~k_2=n_2,~k_3=n_3,~k_4=n_4,~k_5=n_5)\\
& = p_{p_1}(k=n_1, 5)\\ 
& \times p_{\frac{p_2-p_1}{1-p_1}}(k=n_2-n_1,5-n_1) \\
& \times p_{\frac{p_3-p_2}{1-p_2}}(k=n_3-n_2,5-n_2) \\
& \times p_{\frac{p_4-p_3}{1-p_3}}(k=n_4-n_3,5-n_3) \\
& \times p_{\frac{p_5-p_4}{1-p_4}}(k=n_5-n_4,5-n_4).
\end{split}
\end{equation}

The quantity $p_\text{FEAD}$ is the probability that the IEAD probabilities are below the respective values obtained from analysing CDTGs, when obtained by randomly drawing samples from the individual binomial distributions.

From this calculation, we obtain FEAD probabilities (shown in the last row of Table \ref{tab7}) of $0.03\%$ for the $0.50$ CDF values, $0.00\%$ for the $0.33$ CDF values, and $0.01\%$ for the $0.25$ CDF values. These results are all dramatically lower than the $50\%$ probability expected randomly. We conclude that the dynamical clustering procedure used in this work produces CDTGs with significantly lower elemental-abundance dispersions than obtained from random selections of same-sized groups from the full RPE sample.

\subsection{Statistical Significance of the Results}

In the previous two subsections, we obtained IEAD probabilities describing the individual results for the numbers of CDTGs with  elemental-abundance dispersions for CDF values, $v$, below 0.50, 0.33, and 0.25, respectively, as well as the FEAD probabilities, reflecting the full set of elemental-abundance dispersions over all three levels.  

We now evaluate the statistical significance of our results for each of the individual elemental-abundance dispersions -- the Global Elemental-Abundance Dispersion (GEAD) probabilities, as well as the resulting Overall Elemental-Abundance Dispersion (OEAD) probability, calculated in order to assign a single probability for our abundance-dispersion results, and defined as described below. 

For this purpose, we need to assess the probability of finding the triplets of observed numbers summarized in Table \ref{tab7} for the three different CDF values, by randomly drawing values from the binomial distribution. Using the definition for binomial probability $p_\nu(k=n,~N)$ from Section 4, the probability of obtaining exactly $N_1$, $N_2$, and $N_3$ CDTGs out of $N$ with CDF values (for a given elemental-abundance dispersion) below $0.50$, $0.33$, and $0.25$, respectively, is:
\begin{equation}\label{eq6}
\begin{split}
&p(k_1=N_1,~k_2=N_2,~k_3=N_3,~N)\\
&~~~=p_{0.25}(k=N_1,~N) \times p_{\frac{0.33-0.25}{1-0.25}}(k=N_2-N_1,~N-N_1)\\
&~~~\times p_{\frac{0.50-0.33}{1-0.33}}(k=N_3-N_2,~N-N_2),
\end{split}
\end{equation}
and the probability of obtaining a triplet of these or larger numbers of CDTGs below 0.50, 0.33, and 0.25 can be calculated through the conditional sum\footnote{The detailed derivation of these equations can be found in  the Appendix.}:
\begin{equation}\label{eq7}
\begin{split}
&p(k_1\geq N_1,~k_2\geq N_2,~k_3\geq N_3,~N) \\ 
&~~~~~~~=\sum\limits_{\mathclap{\substack{n_1,n_2,n_3\in\mathbb{Z},~~~~~~~~~\\n_1\geq N_1,\\n_2\geq N_2,\\n_3\geq N_3,\\0\leq n_1\leq n_2\leq n_3\leq N~~~~~~~~~~~}}}p(k_1=n_1,k_2=n_2,~k_3=n_3,~N).
\end{split}
\end{equation}
Calculating this sum for the numbers from the third column of Table \ref{tab7} yields the results for the GEAD probabilities, shown in the 5th column of that table. All of these probabilities are substantially below $50\%$ -- in fact, all but one of them are $2\%$ or less, and two of them are below $1\%$. This indicates strong statistical significance of the elemental-abundance dispersions in our CDTGs.  We note that the GEAD probability for [Eu/Fe] (22.2\%), though clearly statistically significant, is large compared to the GEAD probabilities of the other abundance dispersions.  From inspection of the table, this is driven primarily by the larger IEAD probabilities for the CDF levels 0.33 and 0.25, relative to the other abundance dispersions. As an experiment, we dropped the five CDTGs with the lowest CL values (retaining only those CDTGs with CL levels greater than 60\%; the GEAD probability for [Eu/Fe] was reduced to 18.7\%.   

Finally, to obtain a single number characterizing the statistical significance of all of our results (i.e., the probability that they could have been obtained by forming clusters from randomly drawn groups of stars from the RPE Sample), we apply Eqn.~\ref{eq4} listed in Section 5.2 to these five probabilities.  The formal result is the Overall Elemental-Abundance Dispersion (OEAD) probability, $p_{\rm OEAD} << 0.001\%$. This extremely low probability value strongly supports the assertion that clustering on the dynamical quantities produces CDTGs with significantly lower elemental-abundance dispersions than random selection.

We note that, if the hypothesis that the RPE stars were born in dwarf galaxy-like environments holds true, we can go one step further. The [Fe/H] and [C/Fe]$_c$ abundance ratios can be thought of as tracers of the birth environments, while [Sr/Fe], [Ba/Fe], and [Eu/Fe] can be thought of as tracers of the nature of the $r$-process progenitor(s).  In contrast to the above, where all five abundances are considered together for calculation of the OEAD probability, we can group on these two sets of abundance dispersions; the ``environmental variables" (EVs) and the ``progenitor variables" (PVs), respectively. The OEAD probabilities obtained for the EVs (0.02\%) and PVs (0.02\%) are both extremely low, supporting both the birth-environment hypothesis and the universality of the $r$-process signature for the main $r$-process. 

It is worth noting that the estimation of binomial probabilities for groups of elements (e.g., the FEAD, OEAD, EV, and PV probabilities) assumes that their individual abundances are independent. While that is certainly true for their measurement, there are physical reasons to believe that the 
heavy-element abundances ([Sr/Fe], [Ba/Fe], and [Eu/Fe]) may be somewhat correlated with each other. Indeed, the processes that produced sufficient Eu to enrich the stars in a given environment to the point of becoming $r$-process enhanced would also be expected to produce large quantities of Sr and Ba. Although this possible correlation was not taken into account in the present work, we don't expect its presence to significantly affect the results. Formally, the results are independent of this possible correlation -- high statistical significance in the case of independent abundances implies high statistical significance in the case of correlated abundances, and vice versa.

One interesting feature is the clearly less statistically significant [Eu/Fe] abundance dispersions for our CDTGs compared to the other 
heavy elements ([Sr/Fe] and [Ba/Fe]).  We found this result to be 
robust when using other clustering methods, such as FoF, and when changing the \texttt{HDBSCAN} parameters, suggesting it may be real. This result was also found in the previous work of \citet{roederer2018a}. From the observational standpoint, Eu measurements are more challenging than either Sr or Ba, since the former is often based on fewer, weaker lines. Thus it might be expected that the Eu dispersions within CDTGs could be less consistent star-to-star for this reason alone. Larger, homogeneously analyzed, data samples with higher signal-to-noise [Eu/Fe] measurements will be required to test this hypothesis.

\section{Discussion}

\subsection{Membership of CDTGs in Known Substructures, DTGs, or Groups}

We expect that many of the CDTGs we identify in this paper may be members of known substructures, and the DTGs identified by \citet{yuan2020} and \citet{limberg2020}, or the RPE Groups identified by \citet{roederer2018a}.

Table~\ref{tab8} lists the dynamical properties of the known substructures and previously identified DTGs and Groups. The first column lists the substructure, DTG, or Group, along with the numbers of member stars from the RPE Sample it may be associated with. The second column lists the CDTGs which may be associated with each of these.  The final columns list the average dynamical parameters (and dispersions in these quantities) for the CDTGs listed in the second column. We note that the substructures and the DTGs from \citet{yuan2020} and \citet{limberg2020} are of immediate interest, as they were identified without knowledge of the detailed chemistry of their stars. Given that these appear to harbor one or more RPE stars from our sample, it should prove illuminating to obtain high-resolution spectroscopic follow-up of their member stars. Matches with the Groups A-F and Group H of \citet{roederer2018a} are satisfying (since very different clustering algorithms were used), but unsurprising, given that their stars were selected to be RPE stars.

We also note, for completeness, that a number of the DTGs identified in Table~\ref{tab8} were identified in both \citet{yuan2020} and \citet{limberg2020}, but we have chosen to simply list one match. Ultimately, as ongoing and near-future investigations of DTGs and CDTGs are obtained, it will be useful to assemble a catalog of all such identifications, and their constituent stars, which we defer to a future effort.

Regarding associations with better known, prominent substructures previously identified in the halo, over 20\% of the stars in the RPE Sample, and ten of our CDTGs, are apparently associated with \textit{Gaia}-Sausage/Enceladus (GSE) \citep{belokurov2018, helmi2018}, based on their almost null net rotation ($\langle V_{\phi} \rangle \approx -4$\,km\,s$^{-1}$) and high average eccentricities ($\langle e \rangle > 0.8$). The presence of member stars from these CDTGs (see Table \ref{tab8}) in VMP DTGs from \citet{yuan2020} and \citet{limberg2020} corroborate this connection. The progenitor(s) of GSE is/are apparently a major contributor of RPE stars to the inner halo, confirming suggestions from both \citet{yuan2020} and \citet{limberg2020} based on their own DTG analyses.

Two other larger substructures have been recognized in the halo, both predominantly retrograde: Sequoia \citep{myeong2019} and Thamnos \citep{koppelman2019}. Four of our CDTGs could be associated with them (Table \ref{tab8}) according to the independently proposed criteria of \citet{naidu2020}. On the other hand, neither \citet{yuan2020} nor \citet{limberg2020} presented statistical evidence of associations between our studied RPE stars and their DTGs. It is clear that we are in urgent need of high-resolution abundance analyses of stars from Sequoia/Thamnos, particularly for those inhabiting the VMP end of their metallicity distributions, in order to understand the chemical evolution and early star-forming environments of their progenitor(s).

Three member stars from CDTG-15 ($N = 6$, CL $= 83.2\%$) are associated with GL20:DTG-3 \citep{limberg2020}, and one of them (BD$+$30:2611; \citealt{burris2000}) with ZY20:DTG-3 \citep{yuan2020} as well. Both of these DTGs have been attributed to the \citet{helmi1999} stream by these authors. \citet{limberg2020} has also suggested a connection between this stream and RPE stars based on their dynamics. It is striking that spectroscopic studies of this structure had already claimed that its stars were enriched in neutron-capture elements, primarily via the $r$-process, based on  
the close conformity among the abundances of Ba and many elements in the 
rare-earth domain to the scaled-solar $r$-process pattern \citep{roederer2010} or their [Sr/Ba] profiles \citep{aguado2020}. 

The metallicity values of stars from CDTG-15 are in the range $-2.7 \lesssim$ [Fe/H] $\lesssim -1.4$; its average metallicity is $\langle \rm[Fe/H] \rangle = -2.0$, consistent with \citet{aguado2020} and others in the literature (e.g., \citealt{myeongStreamsAndClumps, myeong2018Shards}). Additionally, the [Fe/H] dispersion of $\approx 0.5$ dex found for CDTG-15 leads to a CDF value of 0.71 (commensurate with random draws from the full RPE Sample), consistent with its progenitor being a dwarf galaxy with extended star formation (see \citealt{koppelmanHelmi}).

Finally, there are three CDTGs with similar dynamics to the metal-weak thick disk (MWTD) (Table \ref{tab8}). This component was originally proposed by \citet{morrison1990}, and recently shown to be an independent structure from the canonical thick disk by \citet{carollo2019} and \citet{an2020}. This finding is particularly intriguing, given the recent demonstration that the disk system of the Galaxy harbors significant populations of extremely and ultra metal-poor stars \citep{sestito2019, sestito2020}. These recent efforts, alongside our RPE Sample, point to a MWTD with a diverse chemical profile. Clearly, larger samples of low-metallicity stars with disk-like kinematics will be necessary to unveil the early nucleosynthesis processes that contributed to these peculiar elemental-abundance patterns.

\subsection{Apocentric and Pericentric Distances of the  RPE Sample}

\citet{roederer2018a} noted that only 20\% of the RPE stars they considered have apocentric distances (or $Z_{\rm max}$ distances) greater than 20 kpc. For our much larger sample of RPE stars, we obtain similar results; only 14\% have apocentric distances (or $Z_{\rm max}$ distances) greater than 20 kpc.  These authors pointed out that the present distances to the known UFD or canonical dSph satellite galaxies are all greater than 20 kpc. The limited orbital information presently available for these galaxies (e.g., \citealt{simon2018}) indicates that only the UFD galaxy Tucana III passes within 15 kpc of the Galactic center; the rest all have pericentric distances greater than 20 kpc.  From inspection of the top panel of Figure~\ref{distances_paper}, all of the stars in the RPE Sample have pericentric distances $r_{\rm peri} < 20$ kpc. Many have pericentric distances that pass quite close to the Galactic center: 52\% come within 2.5 kpc, and 26\% come within 1 kpc.

The obvious implication of the above is that RPE stars once associated with parent satellite galaxies that passed too close to the Galactic center were disrupted, and strewn throughout the inner-halo region of the MW, as previously emphasized by \citet{roederer2018a}.   

\subsection{Implications of CDTG Chemistry for the Birth Environments of RPE Stars}

In Section 5, we have described at length the behavior of a limited set of available elemental-abundance ratios ([Fe/H], [C/Fe]$_c$, [Sr/Fe], [Ba/Fe], and [Eu/Fe]) for the 30 CDTGs identified in this work. We have also defined the various probabilities that can be obtained to assess whether or not the observed relatively low dispersions of these abundances are consistent with random draws from the full set of stars in the RPE Sample. The conclusion appears unavoidable that, in fact, they are not. The members of the CDTGs identified on the basis of the similarity in their dynamical parameters shared common chemical-evolution histories, strongly suggesting that they originated in parent satellite galaxies (or globular clusters), which were later disrupted into the inner-halo region of the Galaxy. Most of those parent satellite galaxies were probably not very massive. The average metallicities of most (22/30) of the CDTGs listed in Table~\ref{tab5_stub} are very low ([Fe/H] $\leq -2$), and the galaxy mass-metallicity relationship \citep{kirby2013} implies that they were likely born in relatively low-mass dwarf galaxies, consistent with the findings of \citet{roederer2018a}.

A related conclusion is reached that, although multiple astrophysical progenitors of the $r$-process are not strictly precluded by the present data alone, the evidence in hand is fully consistent with a universal $r$-process pattern, at least for the main $r$-process. Comparison with predictions from high-resolution, large-scale cosmological numerical simulations that consider multiple classes of $r$-process progenitors is required to further explore this hypothesis.

We note the possibility that at least a few of the CDTGs we have identified may have been born in globular clusters, rather than low-mass galaxies, on the sole basis of their small dispersions in metallicity. There are five CDTGs listed in Table~\ref{tab5_stub} with [Fe/H] dispersions of 0.20 dex or less:  CDTG-11 ($-2.43 \pm 0.10$; $N$ = 6, CL = $87.1\%$), CDTG-18 ($-2.12 \pm 0.13$; $N$ = 4, CL = $79.3\%$), CDTG-19 ($-2.48 \pm 0.17$; $N$ = 4, CL = $78.0\%$), CDTG-25 ($-1.93 \pm 0.15$; $N$ = 3, CL = $61.2\%$) and CDTG-27 ($-2.11 \pm 0.08$; $N$ = 3, CL = $37.4\%$), where the numbers in parentheses refer to the mean [Fe/H], dispersion in [Fe/H], number of member stars, and the CDTG confidence level, respectively. It is beyond the scope of this paper to investigate this possibility further, but these CDTGs would be of particular interest for further study in order to accept or reject such associations.

\section{Summary and Conclusions}

We have used accurate dynamical parameters, based primarily on astrometry (parallaxes and proper motions) and radial velocities from \textit{Gaia} DR2 (supplemented by a number of other sources), for a large sample of 446 $r$-process-enhanced stars (RPE: 127 $r$-II and 319 $r$-I stars) assembled by the $R$-Process Alliance (and other sources), to identify 30 Chemo-Dynamically Tagged Groups (CDTGs) in the halo of the Milky Way. A statistical analysis of the dispersions in their chemical abundances for a restricted set of elements ([Fe/H], [C/Fe], [Sr/Fe], [Ba/Fe], and [Eu/Fe]) demonstrates that the members of individual CDTGs have likely experienced common chemical-evolution histories, presumably in their parent satellite galaxies (or, in a few cases, globular clusters) prior to being dispersed into the inner-halo region of the Galaxy. Their observed apocentric and pericentric distances are consistent with this interpretation, similar to the inference drawn by \citet{roederer2018a} based on a much smaller sample. A number of our CDTGs are associated with previously identified substructures, DTGs, and RPE groups.

We note that we have not yet explicitly considered the dynamical clustering of the limited-$r$ stars, since they still remain relatively few in number. This limitation will be lifted in the near future, based on the expected doubling or tripling of the numbers of these, and other RPE stars, once the ongoing analysis of existing and soon-to-be-acquired snapshot high-resolution spectroscopic follow-up from the RPA is completed. Further enlargement of the numbers of RPE stars will also enable the use of more strict dynamical clustering criteria (and hence consideration of clusters with higher confidence levels), and provide for the identification of numerous new CDTGs. Stars in the DTGs identified from lower-resolution spectroscopy that are associated with the RPE CDTGs we have identified are of particular interest as well, as they most likely harbor additional RPE stars that have yet to be recognized. 

We believe that the statistical techniques we have developed in the process of the present analysis for evaluating the significance of the elemental-abundance dispersions in CDTGs will prove useful for future similar observational studies, and for comparison with predictions obtained by chemical-evolution models. Both such efforts are necessary in order to tell the full story of the nature of the astrophysical origin of the $r$-process, in particular the hypothesis of a single predominant astrophysical source suggested by the significantly low elmental-abundance dispersions within our CDTGs.

\acknowledgements

The authors are thankful to an anonymous referee, who provided comments and suggestions that substantially improved this paper.  D.G., D.S., T.C.B., I.U.R., V.M.P., E.M.H., S.D., K.C.R., R.E., and A.F. acknowledge partial support for this work from grant PHY 14-30152; Physics Frontier Center/JINA Center for the Evolution of the Elements (JINA-CEE), awarded by the US National Science Foundation. I.U.R. acknowledges support from grant AST-1815403 awarded by the NSF. G.L. acknowledges CAPES (PROEX; Proc. 88887.481172/2020-00). T.T.H. acknowledges generous support from the George P. and Cynthia Woods Institute for Fundamental Physics and Astronomy at Texas A\&M University.

\clearpage

\begin{deluxetable*}{lrcccrccccc}
	\tabletypesize{\footnotesize}
	\tablecaption{Average Energies, Eccentricities, Actions, Rotational Velocities, and Elemental Abundances for the CDTGs\label{tab6}}
	\tablehead{\colhead{Name} 
		& \colhead{$N$ stars} 
		& \colhead{$\langle Energy \rangle$} 
		& \colhead{$\langle J_r \rangle$ / $\langle J_\phi \rangle$ / $\langle J_z \rangle$} 
		& \colhead{$\langle {\rm Ecc} \rangle$} 
		& \colhead{$\langle V_\phi \rangle $} 
		& \colhead{$\langle {\rm [Fe/H]} \rangle $} 
		& \colhead{$\langle {\rm [C/Fe]}_{c} \rangle $} 
		& \colhead{$\langle {\rm [Sr/Fe]} \rangle$} 
		& \colhead{$\langle {\rm [Ba/Fe]} \rangle$} 
		& \colhead{$\langle {\rm [Eu/Fe]} \rangle$}\\
		\colhead{} & \colhead{} & \colhead{$\times 10^5$} & \colhead{$\times 10^3$} & \colhead{} & \colhead{} & \colhead{} & \colhead{} & \colhead{} & \colhead{} & \colhead{}\\
		\colhead{} & \colhead{} & \colhead{(km$^2$ s$^{-2}$)} & \colhead{(kpc km s$^{-1}$)} & \colhead{} & \colhead{(km s$^{-1}$)} & \colhead{} & \colhead{} & \colhead{} & \colhead{} & \colhead{}}
	\startdata
	CDTG-1  & 6  & $-$1.84 & 0.59 / \hphantom{$-$}0.12 / 0.03 & 0.94 & 15.37        & $-$2.19$\pm$0.28 & $-$0.01$\pm$0.28 & $+$0.08$\pm$0.19 & $+$0.12$\pm$0.23 & $+$0.57$\pm$0.24 \\
	CDTG-2  & 8  & $-$1.82 & 0.05 / $-$0.91            / 0.10 & 0.27 & $-$160.12    & $-$2.35$\pm$0.35 & $+$0.34$\pm$0.29 & $+$0.29$\pm$0.28 & $-$0.05$\pm$0.27 & $+$0.51$\pm$0.24 \\
	CDTG-3  & 3  & $-$1.64 & 0.16 / \hphantom{$-$}1.37 / 0.03 & 0.37 & 159.57       & $-$1.40$\pm$0.28 & $-$0.07$\pm$0.12 & $+$0.48$\pm$0.12 & $+$0.38$\pm$0.27 & $+$0.63$\pm$0.08 \\
	CDTG-4  & 4  & $-$1.20 & 0.69 / \hphantom{$-$}0.55 / 1.91 & 0.51 & 66.69        & $-$2.19$\pm$0.32 & $+$0.15$\pm$0.07 & $+$0.48$\pm$0.34 & $-$0.03$\pm$0.27 & $+$0.48$\pm$0.09 \\
	CDTG-5  & 3  & $-$1.70 & 0.09 / \hphantom{$-$}1.32 / 0.01 & 0.30 & 168.90       & $-$1.75$\pm$0.54 & \dots            & \dots            & $-$0.07$\pm$0.08 & $+$0.45$\pm$0.14 \\
	CDTG-6  & 3  & $-$1.68 & 0.25 / \hphantom{$-$}0.80 / 0.25 & 0.52 & 97.24        & $-$2.20$\pm$0.67 & \dots            & \dots            & $+$0.28$\pm$0.34 & $+$0.80$\pm$0.34 \\
	CDTG-7  & 7  & $-$1.81 & 0.38 / \hphantom{$-$}0.17 / 0.28 & 0.83 & 22.14        & $-$2.49$\pm$0.36 & $+$0.50$\pm$0.15 & $+$0.12$\pm$0.14 & $-$0.03$\pm$0.13 & $+$0.57$\pm$0.19 \\
	CDTG-8  & 3  & $-$1.59 & 0.09 / \hphantom{$-$}1.27 / 0.29 & 0.28 & 150.90       & $-$2.43$\pm$0.31 & $+$0.17$\pm$0.19 & $+$0.21$\pm$0.60 & $-$0.11$\pm$0.06 & $+$0.56$\pm$0.12 \\
	CDTG-9  & 5  & $-$1.74 & 0.77 / \hphantom{$-$}0.00 / 0.01 & 0.95 & 0.23         & $-$1.56$\pm$0.24 & $+$0.14$\pm$0.09 & $+$0.10$\pm$0.13 & $+$0.07$\pm$0.11 & $+$0.46$\pm$0.02 \\
	CDTG-10 & 4  & $-$1.61 & 1.02 / $-$0.07            / 0.00 & 0.98 & $-$8.90      & $-$1.93$\pm$0.41 & $-$0.11$\pm$0.17 & $+$0.20$\pm$0.14 & $+$0.21$\pm$0.31 & $+$0.57$\pm$0.20 \\
	CDTG-11 & 6  & $-$1.68 & 0.46 / \hphantom{$-$}0.67 / 0.10 & 0.70 & 78.72        & $-$2.43$\pm$0.10 & $+$0.20$\pm$0.24 & $-$0.16$\pm$0.27 & $-$0.08$\pm$0.21 & $+$0.59$\pm$0.13 \\
	CDTG-12 & 6  & $-$1.80 & 0.46 / $-$0.35            / 0.07 & 0.80 & $-$46.57     & $-$2.06$\pm$0.24 & $+$0.09$\pm$0.10 & $+$0.22$\pm$0.24 & $-$0.01$\pm$0.35 & $+$0.44$\pm$0.15 \\
	CDTG-13 & 9  & $-$1.69 & 0.72 / $-$0.10            / 0.13 & 0.93 & $-$10.53     & $-$2.20$\pm$0.21 & $+$0.22$\pm$0.10 & $+$0.17$\pm$0.15 & $+$0.11$\pm$0.24 & $+$0.53$\pm$0.18 \\
	CDTG-14 & 12 & $-$1.65 & 0.28 / $-$0.42            / 0.61 & 0.57 & $-$61.11     & $-$2.39$\pm$0.38 & $+$0.20$\pm$0.29 & $+$0.09$\pm$0.23 & $-$0.13$\pm$0.30 & $+$0.54$\pm$0.18 \\
	CDTG-15 & 6  & $-$1.34 & 0.37 / \hphantom{$-$}1.25 / 1.08 & 0.42 & 148.70       & $-$2.00$\pm$0.49 & $+$0.15$\pm$0.51 & $+$0.10$\pm$0.33 & $+$0.03$\pm$0.08 & $+$0.44$\pm$0.06 \\
	CDTG-16 & 8  & $-$1.68 & 0.20 / \hphantom{$-$}0.98 / 0.16 & 0.46 & 127.04       & $-$2.38$\pm$0.38 & $+$0.34$\pm$0.32 & $+$0.13$\pm$0.21 & $+$0.16$\pm$0.20 & $+$0.55$\pm$0.30 \\
	CDTG-17 & 3  & $-$1.28 & 1.85 / \hphantom{$-$}0.08 / 0.11 & 0.97 & 10.33        & $-$1.60$\pm$0.48 & \dots            & \dots            & $+$0.27$\pm$0.51 & $+$0.76$\pm$0.30 \\
	CDTG-18 & 4  & $-$1.55 & 0.39 / $-$0.09            / 0.99 & 0.67 & $-$15.41     & $-$2.12$\pm$0.13 & $+$0.44$\pm$0.03 & \dots            & $-$0.16$\pm$0.11 & $+$0.37$\pm$0.02 \\
	CDTG-19 & 4  & $-$1.85 & 0.40 / $-$0.16            / 0.19 & 0.85 & $-$24.40     & $-$2.48$\pm$0.17 & $+$0.23$\pm$0.16 & $+$0.20$\pm$0.12 & $-$0.07$\pm$0.01 & $+$0.52$\pm$0.12 \\
	CDTG-20 & 5  & $-$1.37 & 1.50 / $-$0.48            / 0.01 & 0.91 & $-$57.73     & $-$2.00$\pm$0.48 & $+$0.05$\pm$0.09 & $+$0.05$\pm$0.17 & $+$0.25$\pm$0.06 & $+$0.54$\pm$0.07 \\
	CDTG-21 & 4  & $-$1.50 & 1.17 / \hphantom{$-$}0.07 / 0.12 & 0.96 & 8.89         & $-$1.90$\pm$0.22 & $+$0.34$\pm$0.02 & $+$0.30$\pm$0.25 & $+$0.06$\pm$0.23 & $+$0.58$\pm$0.12 \\
	CDTG-22 & 4  & $-$1.76 & 0.53 / \hphantom{$-$}0.38 / 0.05 & 0.80 & 49.73        & $-$2.26$\pm$0.31 & $+$0.31$\pm$0.14 & $-$0.03$\pm$0.11 & $-$0.02$\pm$0.13 & $+$0.58$\pm$0.21 \\
	CDTG-23 & 4  & $-$1.78 & 0.22 / \hphantom{$-$}0.88 / 0.03 & 0.50 & 114.55       & $-$2.05$\pm$0.44 & $+$0.05$\pm$0.19 & $+$0.28$\pm$0.12 & $+$0.14$\pm$0.11 & $+$0.59$\pm$0.22 \\
	CDTG-24 & 4  & $-$1.68 & 0.34 / \hphantom{$-$}0.91 / 0.05 & 0.58 & 108.32       & $-$2.38$\pm$0.26 & $+$0.07$\pm$0.24 & $+$0.25$\pm$0.09 & $+$0.16$\pm$0.16 & $+$0.61$\pm$0.16 \\
	CDTG-25 & 3  & $-$1.68 & 0.23 / \hphantom{$-$}1.08 / 0.07 & 0.48 & 137.98       & $-$1.93$\pm$0.15 & $+$0.29$\pm$0.08 & $+$0.15$\pm$0.19 & $+$0.02$\pm$0.15 & $+$0.47$\pm$0.09 \\
	CDTG-26 & 3  & $-$1.21 & 1.35 / $-$1.04            / 0.64 & 0.77 & $-$144.50    & $-$2.26$\pm$0.52 & \dots            & $+$0.07$\pm$0.12 & $-$0.09$\pm$0.24 & $+$0.59$\pm$0.06 \\
	CDTG-27 & 3  & $-$1.61 & 0.12 / $-$1.17            / 0.30 & 0.31 & $-$145.53    & $-$2.11$\pm$0.08 & $+$0.23$\pm$0.20 & $-$0.06$\pm$0.09 & $-$0.08$\pm$0.14 & $+$0.54$\pm$0.28 \\
	CDTG-28 & 4  & $-$1.42 & 0.86 / $-$0.60            / 0.51 & 0.75 & $-$83.92     & $-$2.54$\pm$0.22 & $+$0.46$\pm$0.18 & $+$0.03$\pm$0.25 & $-$0.17$\pm$0.12 & $+$0.48$\pm$0.02 \\
	CDTG-29 & 5  & $-$2.03 & 0.13 / \hphantom{$-$}0.03 / 0.29 & 0.79 & 22.40        & $-$1.84$\pm$0.36 & $-$0.09$\pm$0.12 & $-$0.02$\pm$0.59 & $+$0.23$\pm$0.38 & $+$0.83$\pm$0.34 \\
	CDTG-30 & 3  & $-$1.15 & 1.42 / $-$1.72            / 0.47 & 0.71 & $-$281.50    & $-$2.46$\pm$0.51 & $-$0.01$\pm$0.06 & $+$0.21$\pm$0.51 & $+$0.21$\pm$0.37 & $+$0.87$\pm$0.59
	\enddata 
	
	\tablecomments{For each of the listed elements, the table entries indicate the means and standard deviations (dispersions) of the abundances, when at least 3 measurements of that element were available for a given CDTG.  When the number of available measured elemental abundances is 4 or more, biweight estimates of these quantities are reported,  in order to decrease the influence of potential outliers.}
	
\end{deluxetable*}

\renewcommand{\arraystretch}{1.0}
\setlength{\tabcolsep}{0.5em}

\begin{deluxetable*}{llccccccc}
	\tabletypesize{\footnotesize}
	\tablecaption{CDTGs Identified from the \texttt{HDBSCAN} Algorithm 
		\label{tab5_stub}}
	\tablehead{
		\colhead{Name} & \colhead{Class} & \colhead{[Fe/H]} & \colhead{[C/Fe]} & \colhead{[C/Fe]$_c$} & \colhead{[Sr/Fe]} & \colhead{[Ba/Fe]} & \colhead{[Eu/Fe]} & \colhead{[Eu/H]}
	}
	\startdata
	\multicolumn{9}{c}{} \\
	\multicolumn{9}{c}{\textit{CDTG-1} ($N = 6$, CL $= 99.8\%$)}\\
	2MASS J10492192$-$1154173    & $r$-I           & $-$2.18                       & $-$0.42                       & $-$0.41                       & $-$0.06                       & $-$0.16                       & $+$0.33                       & $-$1.85                       \\
	2MASS J20584918$-$0354340    & $r$-I           & $-$2.36                       & 0.00                        & $+$0.40                       & $-$0.24                       & $-$0.09                       & $+$0.36                       & $-$2.00                       \\
	2MASS J12591462$-$7049592    & $r$-I           & $-$2.45                       & $-$0.60                       & $+$0.16                       & $+$0.25                       & $+$0.25                       & $+$0.50                       & $-$1.95                       \\
	2MASS J11404726$-$0833030    & $r$-I           & $-$1.55                       & $+$0.03                       & $+$0.05                       & $+$0.11                       & $+$0.36                       & $+$0.55                       & $-$1.00                       \\
	2MASS J00405260$-$5122491    & $r$-II          & $-$2.11                       & $-$0.04                       & $-$0.04                       & $+$0.09                       & $-$0.04                       & $+$0.86                       & $-$1.25                       \\
	BD$+$17:3248                 & $r$-II          & $-$2.09                       & $-$0.47                       & $-$0.24                       & $+$0.29                       & $+$0.40                       & $+$0.91                       & $-$1.18                       \\
	$\mu \pm \sigma$ ([X/Y])                 &               & $-$2.19$\pm$0.28 & $-$0.25$\pm$0.27 & $-$0.01$\pm$0.28 & $+$0.08$\pm$0.19 & $+$0.12$\pm$0.23 & $+$0.57$\pm$0.24 & $-$1.54$\pm$0.43 \\
	Cumulative fraction value &               & 0.20                        &                             & 0.70                        & 0.32                        & 0.44                        & 0.68                        &                             \\
	&               &                             &                             &                             &                             &                             &                             &                             \\
	\multicolumn{9}{c}{} \\
	\multicolumn{9}{c}{\textit{CDTG-2} ($N = 8$, CL $= 99.1\%$)}\\
	2MASS J19445483$-$4039459    & $r$-I           & $-$1.98                       & $-$0.27                       & $+$0.33                       & $+$0.21                       & $+$0.13                       & $+$0.33                       & $-$1.65                       \\
	2MASS J15293404$-$3906241    & $r$-I           & $-$2.74                       & \dots                         & \dots                         & $-$0.36                       & $-$0.16                       & $+$0.34                       & $-$2.40                       \\
	RAVE J151558.3$-$203821      & $r$-I           & $-$2.74                       & $-$0.15                       & $+$0.58                       & $+$0.27                       & $-$0.39                       & $+$0.36                       & $-$2.38                       \\
	2MASS J20303339$-$2519500    & $r$-I           & $-$2.21                       & $-$0.65                       & $+$0.10                       & $+$0.41                       & $-$0.29                       & $+$0.41                       & $-$1.80                       \\
	2MASS J15133549$-$1244339    & $r$-I / CEMP-$r$  & $-$2.04                       & $+$0.45                       & $+$0.75                       & $-$2.87                       & $-$0.10                       & $+$0.55                       & $-$1.49                       \\
	2MASS J14232679$-$2834200    & $r$-I           & $-$1.90                       & $+$0.43                       & $+$0.44                       & $+$0.44                       & $-$0.07                       & $+$0.61                       & $-$1.29                       \\
	BPS CS 22896$-$0154          & $r$-II          & $-$2.69                       & $+$0.27                       & $+$0.28                       & $+$0.54                       & $+$0.51                       & $+$0.86                       & $-$1.83                       \\
	2MASS J14592981$-$3852558*   & $r$-II          & $-$2.48                       & $-$0.95                       & $-$0.17                       & $+$0.23                       & $+$0.13                       & $+$0.93                       & $-$1.55                       \\
	$\mu \pm \sigma$ ([X/Y])                 &               & $-$2.35$\pm$0.35 & $-$0.12$\pm$0.52 & $+$0.34$\pm$0.29 & $+$0.29$\pm$0.28 & $-$0.05$\pm$0.27 & $+$0.51$\pm$0.24 & $-$1.73$\pm$0.40 \\
	Cumulative fraction value &               & 0.26                        &                             & 0.72                        & 0.58                        & 0.51                        & 0.64                        &                             \\
	&               &                             &                             &                             &                             &                             &                             &                             \\
	\multicolumn{9}{c}{} \\
	\multicolumn{9}{c}{\textit{CDTG-3} ($N = 3$, CL $= 97.4\%$)}\\
	2MASS J16024498$-$1521016    & $r$-I           & $-$1.80                       & $-$0.05                       & $-$0.05                       & $+$0.35                       & $+$0.08                       & $+$0.55                       & $-$1.25                       \\
	2MASS J09284944$-$0738585    & $r$-I           & $-$1.22                       & 0.00                        & $+$0.07                       & $+$0.45                       & $+$0.32                       & $+$0.60                       & $-$0.62                       \\
	2MASS J10191573$-$1924464    & $r$-II          & $-$1.19                       & $-$0.32                       & $-$0.23                       & $+$0.63                       & $+$0.73                       & $+$0.75                       & $-$0.44                       \\
	$\mu \pm \sigma$ ([X/Y])                 &               & $-$1.40$\pm$0.28 & $-$0.12$\pm$0.14 & $-$0.07$\pm$0.12 & $+$0.48$\pm$0.12 & $+$0.38$\pm$0.27 & $+$0.63$\pm$0.08 & $-$0.77$\pm$0.35 \\
	Cumulative fraction value &               & 0.40                        &                             & 0.28                        & 0.23                        & 0.67                        & 0.22                        &                             \\
	&               &                             &                             &                             &                             &                             &                             &                             \\
	\multicolumn{9}{c}{} \\
	\multicolumn{9}{c}{\textit{CDTG-4} ($N = 4$, CL $= 93.7\%$)}\\
	2MASS J00413026$-$4058547    & $r$-I           & $-$2.58                       & $-$0.49                       & $+$0.20                       & $+$0.04                       & $-$0.27                       & $+$0.38                       & $-$2.20                       \\
	2MASS J02274104$-$0519230*   & $r$-I           & $-$2.38                       & $-$0.70                       & $+$0.08                       & $+$0.72                       & $-$0.18                       & $+$0.42                       & $-$1.96                       \\
	2MASS J00453930$-$7457294    & $r$-I / CEMP-$r$  & $-$2.00                       & $+$0.93                       & $+$0.98                       & $+$0.83                       & $+$0.37                       & $+$0.55                       & $-$1.45                       \\
	2MASS J00182832$-$3900338    & $r$-I           & $-$1.80                       & $-$0.35                       & $+$0.15                       & $+$0.28                       & $+$0.07                       & $+$0.57                       & $-$1.23                       \\
	$\mu \pm \sigma$ ([X/Y])                 &               & $-$2.19$\pm$0.32 & $-$0.50$\pm$0.28 & $+$0.15$\pm$0.07 & $+$0.48$\pm$0.34 & $-$0.03$\pm$0.27 & $+$0.48$\pm$0.09 & $-$1.71$\pm$0.41 \\
	Cumulative fraction value &               & 0.43                        &                             & 0.10                        & 0.79                        & 0.64                        & 0.24                        &                             \\
	&               &                             &                             &                             &                             &                             &                             &                             \\
	\multicolumn{9}{c}{} \\
	\enddata 
	
	\tablecomments{Single (*) mark indicates the 27 stars
		with relative distance errors in the range  20\%$ < \epsilon < 30$\%. See text for details.}  
	
	\tablecomments{Means ($\mu$), standard deviations ($\sigma$), and cumulative fraction values for the elemental abundances in each CDTG with at least 3 measured abundances are listed.  When the number of available measured elemental abundances  is 4 or more, biweight estimates of these quantities are reported, in order to decrease the influence of potential outliers.}
	
	\tablecomments{This table is a stub; the full table is available in electronic form.}
	
\end{deluxetable*}

\begin{deluxetable*}{cccccc}[t]
	\tabletypesize{\footnotesize}
	\tablecaption{CDTG Elemental-Abundance Statistics\label{tab7}}
	\tablehead{\colhead{Abundance} & \colhead{$\#$ CDTGs} & \colhead{$  N < 0.50$, $0.33$, $0.25$} & \colhead{IEAD Probabilities} & \colhead{GEAD Probabilities} & \colhead{OEAD Probability}}
	\startdata
	\text{[Fe/H]} & $30$ &  $21$, ~$14$, ~$12$ & $\phantom{1}2.1\%$, ~$\phantom{1}8.3\%$, ~$\phantom{1}5.5\%$ & $\phantom{1}0.8\%$ & \hfil\multirow{5}{*}{$<< 0.001\%$}\hfill \\
	\text{[C/Fe]$_\text{c}$} & $26$ & $17$, ~$13$, ~$11$ & $11.5\%$, ~$\phantom{1}8.4\%$, ~$\phantom{1}5.4\%$ & $\phantom{1}1.5\%$ & \\
	\text{[Sr/Fe]} & $26$ & $16$, ~$15$, ~$\phantom{1}9$ & $16.3\%$, ~$\phantom{1}0.8\%$, ~$18.0\%$ & $\phantom{1}0.8\%$ & \\
	\text{[Ba/Fe]} & $30$ & $18$, ~$15$, ~$12$ & $18.1\%$, ~$\phantom{1}4.0\%$, ~$\phantom{1}5.1\%$ & $\phantom{1}2.0\%$ & \\
	\text{[Eu/Fe]} & $30$ & $17$, ~$\phantom{1}9$, ~$\phantom{1}8$ & $29.2\%$, ~$70.1\%$, ~$48.6\%$ & $22.2\%$ & \\
	\hline
	\multicolumn{2}{c}{FEAD Probabilities} & & $0.03\%$, ~$0.00\%$, ~$0.01\%$ & &
	\enddata
	\tablecomments{The Individual Elemental-Abundance Dispersion (IEAD) probabilities represent the 
		binomial probabilities for each element for the levels $v$ =  0.50, 0.33, and 0.25, respectively.
		The Full Elemental-Abundance Dispersion (FEAD) probabilities represent the probabilities
		(across {\it all} elements) for the levels $v$ =  0.50, 0.33, and 0.25, respectively.
		The Global Elemental-Abundance Dispersion (GEAD) probabilities represent the probabilities 
		for the triplet of CDF levels for each element.
		The Overall Elemental-Abundance Dispersion (OEAD) probability represents the probability
		(across {\it all} elements) resulting from random draws from the full CDF.  See text for details.}
\end{deluxetable*}

\renewcommand{\arraystretch}{1.0}
\setlength{\tabcolsep}{0.9em}

\begin{deluxetable*}{c|l|ccr}
	\tabletypesize{\small}
	\tablecaption{Dynamical Properties of Known Substructures, DTGs, and RPE Groups\label{tab8}}
	\tablehead{\colhead{Substructure ($n_\text{sub}$)} & \colhead{Groups} & \colhead{$(\langle V_r \rangle$ , $\langle V_\phi \rangle$ , $\langle V_z \rangle)$} & \colhead{$(\langle J_r \rangle$ , $\langle J_\phi \rangle$ , $\langle J_z \rangle)$} & \colhead{$\langle E \rangle $}\\
		\colhead{} & \colhead{} & \colhead{$\sigma_{V_r}$ , $\sigma_{V_\phi}$ , $\sigma_{V_z}$} & \colhead{$\sigma_{J_r}$ , $\sigma_{J_\phi}$ , $\sigma_{J_z}$} & \colhead{$\sigma_E $}\\
		\colhead{} & \colhead{} & \colhead{} & \colhead{$\times 10^3$} & \colhead{$\times 10^5$}\\
		\colhead{} & \colhead{} & \colhead{} & \colhead{(kpc km s$^{-1}$)} & \colhead{(km$^2$ s$^{-2}$)}}
	\startdata
	GSE (57) & CDTG-1,7,9,10,12,13,17,19,20,21,22 & (24.91, $-$3.85, 4.86)                                                                            & (0.67, $-$0.04, 0.08)                                                                             & $-$1.75                      \\
	&                                    & (178.35, 35.19, 69.69)                                                                          & (0.37, 0.25, 0.09)                                                                              & 0.14                      \\
	\hline
	Sequoia\tablenotemark{a} (6)              & CDTG-26,30                         & ($-$10.51, $-$212.84, 113.44)                                                                       & (1.38, $-$1.47, 0.64)                                                                             & $-$1.21                      \\
	&                                    & (241.61, 98.96, 35.36)                                                                          & (0.24, 0.42, 0.02)                                                                              & 0.04                      \\
	\hline
	Thamnos\tablenotemark{a} (11)       & CDTG-2,27                          & (1.91, $-$157.80, $-$36.46)                                                                         & (0.05, $-$0.92, 0.15)                                                                             & $-$1.79                      \\
	&                                    & (61.88, 18.71, 89.76)                                                                           & (0.03, 0.10, 0.10)                                                                              & 0.10                      \\
	\hline
	MWTD\tablenotemark{b} (9) & CDTG-3,5,8                         & ($-$4.30, 158.81, 9.20)                                                                           & (0.10, $+$1.31, 0.02)                                                                             & $-$1.64                      \\
	&                                    & (63.57, 13.56, 76.15)                                                                           & (0.04, 0.07, 0.02)                                                                              & 0.05                      \\
	\hline
	ZY20:DTG-3\tablenotemark{c}  (1)        & CDTG-15                            & (44.37, 148.70, $-$264.30)                                                                        & (0.37, $+$1.25, 1.08)                                                                             & $-$1.34                      \\
	&                                    & (118.57, 53.84, 63.41)                                                                          & (0.13, 0.25, 0.07)                                                                              & 0.05                      \\
	\hline
	ZY20:DTG-7\tablenotemark{d} (1)        & CDTG-1                             & (27.81, 15.37, 6.98)                                                                            & (0.59, $+$0.12, 0.03)                                                                             & $-$1.84                      \\
	&                                    & (41.86, 4.04, 46.33)                                                                            & (0.03, 0.03, 0.01)                                                                              & 0.03                      \\
	\hline
	ZY20:DTG-16\tablenotemark{d} (1)       & CDTG-12                            & (39.03, $-$46.57, 3.59)                                                                           & (0.46, $-$0.35, 0.07)                                                                             & $-$1.80                      \\
	&                                    & (72.13, 6.14, 50.28)                                                                            & (0.02, 0.05, 0.03)                                                                              & 0.03                      \\
	\hline
	ZY20:DTG-46\tablenotemark{d} (1)       & CDTG-28                            & (38.76, $-$83.92, 8.27)                                                                           & (0.86, $-$0.60, 0.51)                                                                             & $-$1.42                      \\
	&                                    & (246.05, 50.62, 99.95)                                                                          & (0.02, 0.16, 0.06)                                                                              & 0.04                      \\
	\hline
	GL20:DTG-2 (3)        & CDTG-4                             & ($-$91.85, 66.69, $-$22.65)                                                                         & (0.69, $+$0.55, 1.91)                                                                             & $-$1.20                      \\
	&                                    & (33.31, 36.71, 321.86)                                                                          & (0.20, 0.26, 0.11)                                                                              & 0.01                      \\
	\hline
	GL20:DTG-3\tablenotemark{c}  (3)        & CDTG-15                            & (44.37, 148.70, $-$264.30)                                                                        & (0.37, $+$1.25, 1.08)                                                                             & $-$1.34                      \\
	&                                    & (118.57, 53.84, 63.41)                                                                          & (0.13, 0.25, 0.07)                                                                              & 0.05                      \\
	\hline
	GL20:DTG-7 (2)        & CDTG-28                            & (38.76, $-$83.92, 8.27)                                                                           & (0.86, $-$0.60, 0.51)                                                                             & $-$1.42                      \\
	&                                    & (246.05, 50.62, 99.95)                                                                          & (0.02, 0.16, 0.06)                                                                              & 0.04                      \\
	\hline
	GL20:DTG-30\tablenotemark{d} (2)       & CDTG-13                            & ($-$45.29, $-$10.53, 15.73)                                                                         & (0.72, $-$0.10, 0.13)                                                                             & $-$1.69                      \\
	&                                    & (198.01, 19.10, 79.35)                                                                          & (0.07, 0.14, 0.04)                                                                              & 0.01                      \\
	\hline
	IR18:Group A (2)      & CDTG-11,23                         & ($-$23.57, 84.56, 10.25)                                                                          & (0.43, $+$0.73, 0.07)                                                                             & $-$1.71                      \\
	&                                    & (90.30, 20.95, 61.77)                                                                           & (0.14, 0.14, 0.04)                                                                              & 0.05                      \\
	\hline
	IR18:Group B (1)      & CDTG-13                            & ($-$45.29, $-$10.53, 15.73)                                                                         & (0.72, $-$0.10, 0.13)                                                                             & $-$1.69                      \\
	&                                    & (198.01, 19.10, 79.35)                                                                          & (0.07, 0.14, 0.04)                                                                              & 0.01                      \\
	\hline
	IR18:Group C (2)      & CDTG-16                            & (53.27, 127.04, $-$33.27)                                                                         & (0.20, $+$0.98, 0.16)                                                                             & $-$1.68                      \\
	&                                    & (76.88, 23.75, 77.77)                                                                           & (0.02, 0.05, 0.02)                                                                              & 0.02                      \\
	\hline
	IR18:Group D (1)      & CDTG-10                            & ($-$204.81, $-$8.90, 18.44)                                                                         & (1.02, $-$0.07, 0.00)                                                                             & $-$1.61                      \\
	&                                    & (19.37, 11.98, 21.57)                                                                           & (0.04, 0.10, 0.00)                                                                              & 0.02                      \\
	\hline
	IR18:Group E (2)      & CDTG-1                             & (27.81, 15.37, 6.98)                                                                            & (0.59, $+$0.12, 0.03)                                                                             & $-$1.84                      \\
	&                                    & (41.86, 4.04, 46.33)                                                                            & (0.03, 0.03, 0.01)                                                                              & 0.03                      \\
	\hline
	IR18:Group F (1)      & CDTG-22                            & ($-$113.36, 49.73, 8.26)                                                                          & (0.53, $+$0.38, 0.05)                                                                             & $-$1.76                      \\
	&                                    & (69.95, 8.02, 62.15)                                                                            & (0.03, 0.04, 0.02)                                                                              & 0.02                      \\
	\hline
	IR18:Group H (1)      & CDTG-19                            & (9.73, $-$24.40, 39.72)                                                                           & (0.40, $-$0.16, 0.19)                                                                             & $-$1.85                      \\
	&                                    & (82.12, 10.27, 76.64)                                                                           & (0.02, 0.08, 0.02)                                                                              & 0.02                     
	\enddata
	\tablenotetext{a}{Tentative associations with Sequoia \citep{myeong2019} and Thamnos \citep{koppelman2019} substructures.}
	\tablenotetext{b}{Metal-Weak Thick Disk; originally identified by \citet{morrison1990}, and recently shown to be an independent structure from the canonical thick disk by \citet{carollo2019} and \citet{an2020}.}
	
	\tablenotetext{c}{These DTGs have been attributed to the \citet{helmi1999} Stream by \citet{yuan2020} and \citet{limberg2020}.} 
	
	\tablenotetext{d}{These DTGs have been attributed to the \textit{Gaia}-Sausage by \citet{yuan2020} and \citet{limberg2020}.}
	
	\tablecomments{In the first column, the number in parenthesis indicates the number of RPE stars from our study associated with the listed substructure, DTG, or group.}
\end{deluxetable*}

\section{Appendix: Derivation of the Probability Equations}
Here we provide a derivation of the two equations, Eqn. (5) and Eqn. (6),  provided in Section 5.3 for calculation of the Global Elemetal-Abundance Dispersion (GEAD) probabilities from basic principles of Probability Theory.

First, we are interested in the following question: In a sequence of $N$ draws from the uniform distribution over range [0,1], what is the probability that exactly $N_1$ of them will take values below $0.25$, $N_2$ of below $0.33$, and $N_3$ of below $0.50$? In Section 5, we provided the definition of the binomial probability:
\begin{equation}\label{eq8_1}
p_\nu(k=n,~N)=C_N^n\nu^n(1-\nu)^{N-n}
\end{equation}
that, with the probability $p$ of success for an individual experiment, the total number of successes out of $N$ experiments will be $n$. We can apply this probability to our present situation as well.

\begin{figure}[t]
	\includegraphics[width=\columnwidth]{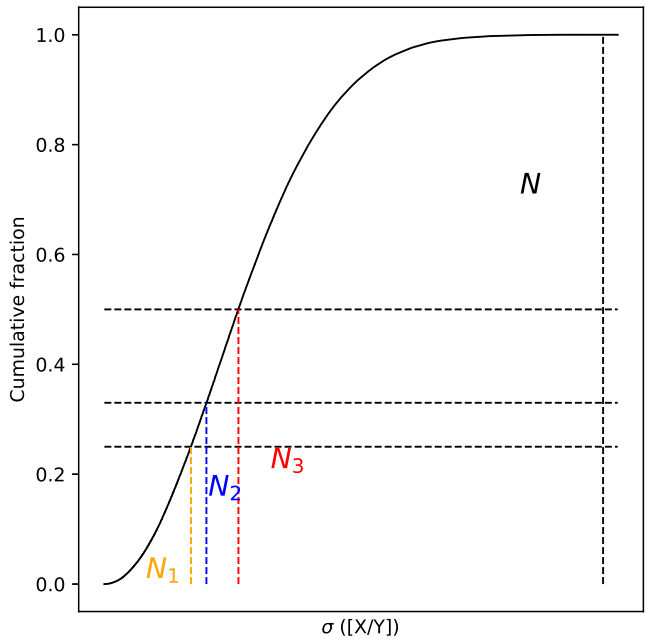}
	\caption{Setup for derivation of the binomial probability. The cumulative fraction distribution of some elemental-abundance dispersion is plotted versus the dispersion value. Here $N_1$ is the number of CDF values below 0.25, $N_2$ below 0.33, and $N_3$ below 0.50, among the total $N$. We are interested in the probability that these numbers can be obtained from a binomial distribution.}
	\label{appendix_cumul_paper}
\end{figure}

Figure~\ref{appendix_cumul_paper} visualizes the problem when applied to cumulative fraction distributions of abundance dispersions. We need to find the probability $p(k_1=N_1,~k_2=N_2,~k_3=N_3,~N)$ that exactly these numbers of CDF values in the specified ranges (below 0.25, below 0.33, and below 0.50) can be obtained by randomly grouping stars from the RPE Sample and calculating the CDF values of the resulting CDTGs.

First, let us find the probability that exactly $N_1$ CDTGs will have CDF values below 0.25. Since the probability of success for each experiment here is exactly 0.25 (as the entire range of CDF values is [0,1], we obtain $0.25/(1-0)=0.25$), hence the first conditional probability will simply equal the first conditional binomial probability:
\begin{equation}\label{eq8_2}
p(k_1=N_1,~N) ~=~ p_{0.25}(k=N_1,~N).
\end{equation}
The next step is to find the probability that, given that exactly $N_1$ values are below $0.25$, that exactly $N_2$ values are below $0.33$. This means that exactly $N_2-N_1$ values must lie in the range [0.25,0.33]. Since all the remaining values are only allowed to fall within the range [0.25,1], the entire range of allowed values is $1-0.25$, while the range of values that must be occupied is $0.33-0.25$. Hence, the probability of one success is $\frac{0.33-0.25}{1-0.25}$, and, given that we have only $N-N_1$ values left to draw from, and exactly $N_2-N_1$ of them must fall within this range, we obtain the second conditional binomial probability:
\begin{equation}\label{eq9}
p(k_2=N_2,~N~|~k_1=N_1) ~=~ p_\frac{0.33-0.25}{1-0.25}(k=N_2-N_1,~N-N_1).
\end{equation}
Similarly, since exactly $N_3-N_2$ values must lie in the range [0.33,0.50], the entire allowed range being $1-0.33$ and the number of values remaining being $N-N_2$, we obtain the third conditional binomial probability:
\begin{equation}\label{eq10}
\begin{split}
p(k_3=N_3,~N~|~k_1=N_1,&~k_2=N_2)~=~\\
&p_\frac{0.50-0.33}{1-0.33}(k=N_3-N_2,~N-N_2).
\end{split}
\end{equation}
The full probability is then found by multiplication of these three conditional probabilities:
\begin{equation}\label{eq11}
\begin{split}
p(k_1=N_1,&~k_2=N_2,~k_3=N_3,~N)~=~\\
&p(k_1=N_1,~N)\times p(k_2=N_2,~N~|~k_1=N_1)\\
&\times p(k_3=N_3,~N~|~k_1=N_1,~k_2=N_2),
\end{split}
\end{equation}
which is Eqn. (5) in Section 5.3.

Now we seek the probability that this, or a higher number of abundance dispersions, fall in these ranges.  That is, the probability that the number of dispersions below 0.25 is equal to or above $N_1$, the number of dispersions below 0.33 is equal to or above $N_2$, and the number of dispersions below 0.50 is equal to or above $N_3$. This can be found by summing over all probabilities with the respective numbers of dispersions $n_1$, $n_2$, and $n_3$ satisfying the conditions $n_1\geq N_1$, $n_2\geq N_2$, $n_3\geq N_3$, and such that $N_1+N_2+N_3\leq N$. This yields the sum:
\begin{equation}\label{eq12}
\begin{split}
&p(k_1\geq N_1,~k_2\geq N_2,~k_3\geq N_3,~N)~=~\\ 
&~~~~~~~\sum\limits_{\mathclap{\substack{n_1,n_2,n_3\in\mathbb{Z},~~~~~~~~~\\n_1\geq N_1,\\n_2\geq N_2,\\n_3\geq N_3,\\0 \leq n_1\leq n_2\leq n_3\leq N~~~~~~~~~~~}}}p(k_1=n_1,k_2=n_2,~k_3=n_3,~N),
\end{split}
\end{equation}
which is Eqn. (6) in Section 5.3 (the condition $n_1,n_2,n_3\in\mathbb{Z}$ denotes that $n_1,n_2,n_3$ must be integers).

\appendix

\section{Full Long Tables}

\renewcommand{\arraystretch}{1.0}
\setlength{\tabcolsep}{0.30em}

\setcounter{table}{1}

\clearpage

\startlongtable
\begin{deluxetable*}{lllrrrrrrcl}
\tabletypesize{\tiny}
\tablecaption{RPA Classes, Temperatures, and Elemental Abundances for the Initial Sample\label{tab3}}
\tablehead{
 \colhead{Name} & \colhead{RPA Class} & \colhead{$T_{\rm eff}$} & \colhead{[Fe/H]} & \colhead{[C/Fe]} & \colhead{[C/Fe]$_c$} & \colhead{[Sr/Fe]} & \colhead{[Ba/Fe]} & \colhead{[Eu/Fe]} & \colhead{Reference} & \colhead{RPA} \\
 \colhead{} & \colhead{} & \colhead{(K)} & \colhead{} & \colhead{} & \colhead{} & \colhead{} & \colhead{} & \colhead{} & \colhead{} & \colhead{}
 }
\startdata
2MASS J00002259$-$1302275      & $r$-I           & 4576 & $-$2.90 & $-$0.65 & $-$0.65 & $-$1.20 & $-$0.38 & $+$0.58 & \textsuperscript {a}   & 1 \\
2MASS J00002416$-$1107454      & $r$-I           & 4693 & $-$2.43 & $-$0.27 & $+$0.30 & $-$0.04 & $-$0.24 & $+$0.51 & \textsuperscript {b}   & 1 \\
SMSS J000113.96$-$363337.9     & limited-$r$     & 4810 & $-$2.32 & $-$0.33 & $+$0.09 & $-$0.04 & $-$1.21 & $-$0.55 & \textsuperscript {c}   & 0 \\
HD 224930                    & $r$-I           & 5275 & $-$1.00 & $\dots$   & $\dots$   & $\dots$   & $-$0.19 & $+$0.34 & \textsuperscript {d}   & 0 \\
2MASS J00021668$-$2453494      & $r$-I           & 5020 & $-$1.81 & $-$0.88 & $-$0.57 & $+$0.59 & $+$0.10 & $+$0.52 & \textsuperscript {a}   & 1 \\
BPS CS 22957$-$0036            & $r$-I           & 5970 & $-$2.28 & $+$0.04 & $+$0.04 & $+$0.20 & $-$0.22 & $+$0.55 & \textsuperscript {e}  & 0 \\
HD 20                        & $r$-II          & 5445 & $-$1.58 & $-$0.34 & $-$0.32 & $+$0.13 & $+$0.32 & $+$0.80 & \textsuperscript {f}   & 0 \\
2MASS J00073817$-$0345509      & $r$-II          & 4663 & $-$2.09 & $-$0.32 & $+$0.17 & $+$0.41 & $+$0.11 & $+$0.73 & \textsuperscript {g}   & 1 \\
\enddata
\end{deluxetable*}

\clearpage

\startlongtable
\begin{deluxetable*}{lcrcccrrr}
\tablecaption{ Derived Dynamical Parameters for the RPE Sample\label{tab4}}
\tablehead{
 \colhead{Name} & \colhead{Energy} & \colhead{Ecc} & \colhead{$J_r$} & \colhead{$J_\phi$}  & \colhead{$J_z$} & \colhead{$r_\text{peri}$} & \colhead{$r_\text{apo}$} & \colhead{$Z_\text{max}$} \\
 \colhead{} &  \colhead{$\times 10^5$} & \colhead{} & \colhead{$\times 10^3$} & \colhead{$\times 10^3$} & \colhead{$\times 10^3$} & \colhead{} & \colhead{} & \colhead{}\\
 \colhead{} & \colhead{(km$^2$ s$^{-2}$)} & \colhead{} & \colhead{(kpc km s$^{-1}$)} & \colhead{(kpc km s$^{-1}$)}  & \colhead{(kpc km s$^{-1}$)} & \colhead{(kpc)} & \colhead{(kpc)} & \colhead{(kpc)}
 }
 \startdata
2MASS J00002259$-$1302275*     & $-$1.50 & 0.79 & 0.74  & $+$0.57 & 0.34 & 1.68  & 14.46   & 9.65   \\
2MASS J00002416$-$1107454*     & $-$1.39 & 0.73 & 0.85  & $-$0.87 & 0.46 & 2.85  & 18.29   & 12.07  \\
HD 224930                    & $-$1.68 & 0.25 & 0.07  & $+$1.41 & 0.01 & 4.86  & 8.13    & 0.42   \\
2MASS J00021668$-$2453494*     & $-$1.42 & 0.40 & 0.25  & $+$0.92 & 1.07 & 6.18  & 14.51   & 12.61  \\
BPS CS 22957$-$0036            & $-$1.72 & 0.84 & 0.58  & $-$0.32 & 0.10 & 0.82  & 9.44    & 4.26   \\
HD 20                        & $-$1.57 & 0.98 & 1.15  & $+$0.06 & 0.01 & 0.15  & 13.44   & 5.68   \\
2MASS J00073817$-$0345509*     & $-$1.71 & 0.86 & 0.75  & $+$0.04 & 0.10 & 0.73  & 10.08   & 3.45   \\
2MASS J00093394$-$1857008      & $-$1.74 & 0.64 & 0.37  & $-$0.74 & 0.04 & 1.96  & 8.96    & 1.37   \\
\enddata 

\tablecomments{Single (*) mark indicates the 16 stars 
with relative distance errors in the range  20\%$ < \epsilon < 30$\%. See text for details.}   
\end{deluxetable*}
\clearpage

\renewcommand{\arraystretch}{1.0}
\setlength{\tabcolsep}{0.30em}

\startlongtable
\begin{deluxetable*}{llccccccc}
\tabletypesize{\tiny}
\tablecaption{CDTGs Identified from the \texttt{HDBSCAN} Algorithm 
\label{tab5}}
\tablehead{
 \colhead{Name} & \colhead{Class} & \colhead{[Fe/H]} & \colhead{[C/Fe]} & \colhead{[C/Fe]$_c$} & \colhead{[Sr/Fe]} & \colhead{[Ba/Fe]} & \colhead{[Eu/Fe]} & \colhead{[Eu/H]}
 }
\startdata
\multicolumn{9}{c}{} \\
\multicolumn{9}{c}{\textit{CDTG-1} ($N = 6$, CL $= 99.8\%$)}\\
2MASS J10492192$-$1154173    & $r$-I           & $-$2.18                       & $-$0.42                       & $-$0.41                       & $-$0.06                       & $-$0.16                       & $+$0.33                       & $-$1.85                       \\
2MASS J20584918$-$0354340    & $r$-I           & $-$2.36                       & 0.00                        & $+$0.40                       & $-$0.24                       & $-$0.09                       & $+$0.36                       & $-$2.00                       \\
2MASS J12591462$-$7049592    & $r$-I           & $-$2.45                       & $-$0.60                       & $+$0.16                       & $+$0.25                       & $+$0.25                       & $+$0.50                       & $-$1.95                       \\
2MASS J11404726$-$0833030    & $r$-I           & $-$1.55                       & $+$0.03                       & $+$0.05                       & $+$0.11                       & $+$0.36                       & $+$0.55                       & $-$1.00                       \\
2MASS J00405260$-$5122491    & $r$-II          & $-$2.11                       & $-$0.04                       & $-$0.04                       & $+$0.09                       & $-$0.04                       & $+$0.86                       & $-$1.25                       \\
BD$+$17:3248                 & $r$-II          & $-$2.09                       & $-$0.47                       & $-$0.24                       & $+$0.29                       & $+$0.40                       & $+$0.91                       & $-$1.18                       \\
\enddata 

\tablecomments{Single (*) mark indicates the 27 stars
with relative distance errors in the range  20\%$ < \epsilon < 30$\%. See text for details.}  

\tablecomments{Means ($\mu$), standard deviations ($\sigma$), and cumulative fraction values for the elemental abundances in each CDTG with at least 3 measured abundances are listed.  When the number of available measured elemental abundances  is 4 or more, biweight estimates of these quantities are reported, in order to decrease the influence of potential outliers.}
\vspace{-0.5cm}
\end{deluxetable*}

\clearpage

\setcounter{table}{0}

\startlongtable
\begin{deluxetable*}{lrrrllrrr}
\tabletypesize{\tiny}
\tablecaption{Positions, Magnitudes, Distances, Radial Velocities, and Proper Motions for the Initial Sample\label{tab2c}}
\tablehead{
\colhead{Name} & \colhead{RA} & \colhead{DEC} & \colhead{$V$ mag} & \colhead{$d_{Gaia}$ (err)} & \colhead{$d_{\rm SH}$ (err)} & \colhead{Rad Vel} & \colhead{PM$_{\rm RA}$} & \colhead{PM$_{\rm DEC}$} \\
\colhead{} & \colhead{(J2000)} & \colhead{(J2000)}  & \colhead{} & \colhead{(kpc)} & \colhead{(kpc)} & \colhead{(km s$^{-1}$)} & \colhead{(mas yr$^{-1}$)} & \colhead{(mas yr$^{-1}$)} 
}
\startdata
2MASS J00002259$-$1302275      & 00:00:22.60 & $-$13:02:27.5 & 12.88 & 5.71 (1.54)*    & 4.78 (1.47)**  & $-$104.4                      & $-$2.61    & $-$9.45    \\
2MASS J00002416$-$1107454      & 00:00:24.16 & $-$11:07:45.4 & 12.04 & 3.76 (0.68)     & 3.51 (0.79)*   & $-$110.4                      & $-$0.63    & $-$23.48   \\
SMSS J000113.96$-$363337.9     & 00:01:13.99 & $-$36:33:38.1 & 13.91 & 8.02 (2.24)*    & 7.28 (2.03)*   & 243.3 \textsuperscript {a}                 & 8.35     & $-$5.16    \\
HD 224930                    & 00:02:11.18 & $+$27:04:40.0 & 5.67  & 0.01 (0.00)     & 0.01 (0.00)    & $-$35.6                       & 723.11   & $-$933.75  \\
2MASS J00021668$-$2453494      & 00:02:16.70 & $-$24:53:49.5 & 13.49 & 4.86 (0.74)     & 5.11 (1.36)*   & 80.1                        & $-$3.61    & $-$8.09    \\
BPS CS 22957$-$0036            & 00:03:31.20 & $-$04:44:19.0 & 14.41 & 1.26 (0.09)     & 1.20 (0.15)    & $-$154.5 \textsuperscript {b}  & 38.64    & $-$22.27   \\
HD 20                        & 00:05:15.48 & $-$27:16:18.8 & 9.01  & 0.50 (0.01)     & 0.50 (0.02)    & $-$56.5                       & 132.43   & $-$39.92   \\
2MASS J00073817$-$0345509      & 00:07:38.17 & $-$03:45:50.9 & 11.65 & 2.86 (0.56)     & 2.72 (0.65)*   & $-$146.3                      & 2.30     & $-$16.04   \\
\enddata
\end{deluxetable*}

{\color{white}This is text to try and force the page. I don't know if this will work or not but at this point I have no idea}

\bibliography{references.bib}

{\color{white}This is text to try and force the page. I don't know if this will work or not but at this point I have no idea}

\end{document}